\documentclass[english]{elsarticle}
\usepackage{ae,aecompl}
\usepackage{helvet}

\usepackage[T1]{fontenc}
\usepackage[latin9]{inputenc}
\usepackage{amsmath}
\usepackage{amssymb}
\usepackage{graphicx}

\makeatletter

\usepackage{babel}

\usepackage{babel}

\makeatother

\usepackage{babel}
\begin{document}

\begin{frontmatter}{}

\title{Microscopic model of quantum butterfly effect: out-of-time-order
correlators and traveling combustion waves}

\author{Igor L. Aleiner }

\address{Columbia Physics Department, 530 W 120th Street, New York, NY, 10027,
USA}

\author{Lara Faoro and Lev B. Ioffe}

\address{Laboratoire de Physique Theorique et Hautes Energies, CNRS UMR 7589 \\ Universites Paris 6 et 7, place Jussieu 75252 Paris, Cedex 05 France}

\address{Department of Physics and Astronomy, Rutgers University, 136 Frelinghuysen
Rd, Piscataway 08854, New Jersey, USA}

\begin{abstract}
We extend the Keldysh technique to enable the computation of out-of-time
order correlators such as $\left\langle O(t)\tilde{O}(0)O(t)\tilde{O}(0)\right\rangle $.
We show that the behavior of these correlators is described by equations
that display initially an exponential instability which is followed
by a linear propagation of the decoherence between two initially identically
copies of the quantum many body systems with interactions. At large
times the decoherence propagation (quantum butterfly effect) is described
by a diffusion equation with non-linear dissipation known in the theory
of combustion waves. The solution of this equation is a propagating
non-linear wave moving with constant velocity despite the diffusive
character of the underlying dynamics.

Our general conclusions are illustrated by the detailed computations
for the specific models describing the electrons interacting with
bosonic degrees of freedom (phonons, two-level-systems etc.) or with
each other. 
\end{abstract}

\end{frontmatter}{}

\tableofcontents{}

\section{Motivation\label{sec:Motivation}}

In a chaotic classical system a small perturbation leads to the exponential
divergence of trajectories characterized by Lyapunov time, $1/\Lambda$.
As a result, the observables in two copies of the system experiencing
different perturbations quickly become uncorrelated. In a many body
system a \emph{local }perturbation initially destroys the correlations
locally, then the region where the correlations are destroyed quickly
grows with time. Killing a butterfly in Ray Bradbury story \cite{Bradbury}
leads to the spreading perturbation until it reaches the size of the
system (Earth in this story). This phenomena is known as butterfly
effect.

The concept of butterfly effect can be generalized to a closed chaotic
quantum system even though such generic system does not necessarily
have a direct analogue of Lyapunov divergence of trajectories because
quantum mechanics prohibits the infinitesimal shift of the trajectory.
The convenient measure of the butterfly effect is provided by the
out-of-time-order correlator (OTOC) that was first introduced by Larkin
and Ovchinnokov\cite{Larkin1969}, revived by Kitaev \cite{Kitaev2014,Kitaev2015}
and extensively discussed by a number of works recently \cite{Maldacena2015,Shenker2014,Roberts2015,Swingle2016}.
OTOC is defined by 
\begin{equation}
\mathcal{A}(t)=\left\langle O(t)\tilde{O}(0)O(t)\tilde{O}(0)\right\rangle ,\label{eq:A_general}
\end{equation}
where $O(t)$ and $\tilde{O}(t)$ are two local operators in Heisenberg
picture. Physically, it describes how much the perturbation introduced
by $\tilde{O}(0)$ changes the value of the $O(t)$. At large times
$\mathcal{A}(t)$ goes to a zero, because the state created by the
consecutive action of the operators $O(t)\tilde{O}(0)$ is incoherent
with the state obtained when these operators act in a different order.\footnote{Here we assume that operators $O$ have zero averages in all states.
If not, the irreducible correlators have to be discussed. We also
assume that operator $\left\langle O^{2}(t)\right\rangle \neq0$ } The anomalous time order in the correlator (\ref{eq:A_general})
implies the evolution backward in time, so it is not measurable by
direct physical experiments on one copy of the system in the absence
of a time machine such as implemented in NMR experiments \cite{Pastawski2016}. One can view the the decrease of the OTOC with time as the consequence
of the dephasing between two initially almost identical Worlds evolving
with the same Hamiltonian. In this respect it is different from the
problems of fidelity \cite{Peres1984} and Loschmidt echo ( \cite{Pastawski2016}
and references therein) that study evolution forward and backwards
with slightly different Hamiltonians. It is also different from a
problem of the evolution of a particle along quasiclassically close
trajectories appearing in studies of the proximity effects \cite{Larkin1969}
or weak-localization \cite{Aleiner1996} and quantum noise \cite{Aleiner2000}. 

For physical systems the Hamiltonian is local, so that distant parts
of a system are not interacting directly with each other. In this
case, one may further distinguish the case when operators $O$ and
$\tilde{O}$ act far from each other in real space. One expects that
the correlator decreases after the significant delay needed for the
perturbation to spread over the distance separating these operators.
When correlators of this type decayed for \emph{any} separation between
the operators in the real space the coherence is completely lost.
The decay of OTOC at long times for all subsystems (i.e. for all separations)
for all operators $O$ and $\tilde{O}$ implies complete quantum information
scrambling \cite{Hosur2016}. Note that the separation of the operators
in space is equivalent to the separation into subsystems introduced
in quantum information works. We are not going to discuss here quantum
information implications of OTOC and the exact definition of quantum
scrambling; we refer the reader to the literature that discussed its
theory \cite{Sekino2008,Lashkari2011,Brown2012,Shenker2013,Page1993}
and the possibility of its experimental measurement \cite{Swingle2016a,Yao2016}. 

The goal of this work is to develop the analytic tools to study OTOC
(\ref{eq:A_general}) for microscopic models that allow for the solutions
for conventional correlators. The technique that we develop is essentially
a straightforward extension of the Keldysh technique. We apply our
technique to three models that are basic in condensed matter physics:
(i) electrons interacting with localized bosonic degrees of freedom
(Einstein phonons or simplified two level systems), (ii) electrons
in the disorder potential and (iii) electrons weakly interacting with
each other. We find that in models (i) and (iii) the mathematical
description of the OTOC is similar to the description of the combustion
waves. The small initial perturbation first grows exponentially remaining
local and then starts to propagate with a constant velocity and a
well defined front, despite the fact that the thermal transport in
these models is always diffusive. The velocity of the front propagation
is always slower than electron Fermi velocity and it is parametrically
slower than it in some models. The apparently slow velocity of the
front propagation implies that it does not necessarily saturate Lieb-Robinson
bound \cite{Roberts2016Lieb-Robinson}. This conclusion of the constant
velocity of the quantum butterfly propagation agrees with the result
obtained in holographical theory of black holes \cite{Shenker2014,Shenker2015,Blake2016}.

The plan of this paper is the following. In Section \ref{sec:Augmented-Keldysh}
we introduce the basic elements of our technique: the augmented Keldysh
formalism that involves two forward and two backward paths. The state
of the system in this formalism is described by the diagonal and off-diagonal
Green functions in the augmented space. The diagonal functions describe
the quasiparticle distribution function in each ``world'', the off-diagonal
ones describe the coherence between the ``worlds''. In Section \ref{subsec:Observables-and-computables}
we introduce two types of correlators that one can compute in this
technique: the observables that can be measured directly in a physical
experiment and the computables that one can only compute numerically
(or measure given the time machine). In Section \ref{sec:Microscopic-Models}
we introduce the details of the microscopic models for which the anomalous
correlator of type (\ref{eq:A_general}) will be computed. In Section
\ref{sec:Kinetic-equation-for} we derive the analogue of the kinetic
equation for both diagonal and off-diagonal ones. In Section \ref{sec:Instability-of-the}
we analyze the stability of the kinetic equations of Section \ref{sec:Kinetic-equation-for}
ignoring their spatial structure (i.e. in zero dimensional case) and
show that the instability of the off-diagonal functions is described
by non-linear ordinary differential equations. Section \ref{sec:Spatial-structure-of}
generalizes these equations for the models with spatial structure
for which they become similar to the equations describing the combustion
waves. Section \ref{sec:Spatial-Propagation-of} describes the formation
of the propagating front that follows from the non-linear diffusive
equations derived in Section \ref{sec:Spatial-structure-of}. The
Section \ref{sec:Initial-conditions-for} studies the initial time
period at which the state of the system is not yet accounted for by
the diffusive equations and its match to the evolution at longer times.
Finally, the Section \ref{sec:Discussion-and-conclusions} gives the
summary of the results and discussions of possible extensions.

\section{Augmented Keldysh and Keldysh techniques\label{sec:Augmented-Keldysh}}

\subsection{Augmented Keldysh technique\label{subsec:Augmented-Keldysh-technique}}

\begin{figure}[h]
\centering\includegraphics[width=0.8\textwidth]{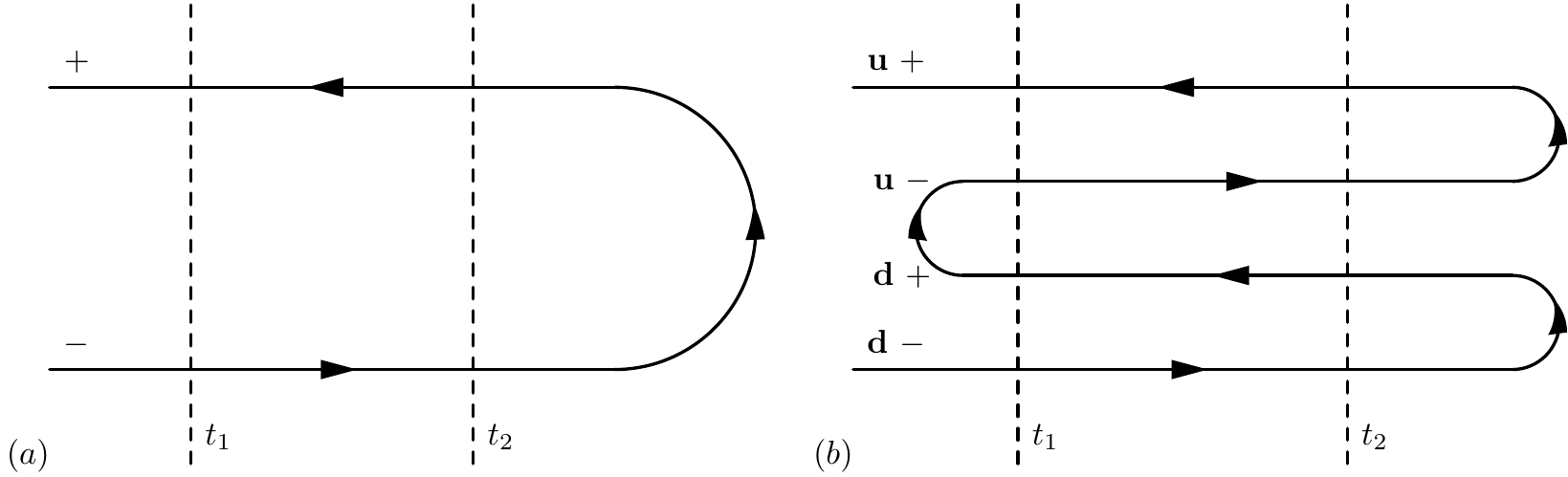}\caption{The traditional, $\mathcal{C}_{K}$, (a) and the augmented (b) Keldysh
contours $\mathcal{C}_{aK}$. Times $t_{_{1,2}}$ label the insertion
of the operators for observable or computable quantities, see text.
Operators (fermionic or bosonic) are ordered according their location
on the contours $\mathcal{C}_{K}$, $\mathcal{C}_{aK}$.\label{fig:The-augmented-Keldysh}}
\end{figure}

Anomalously ordered correlator such as out of time ordered $\mathcal{A}(t)$
introduced in the Section \ref{sec:Motivation}, see Eq. (\ref{eq:A_general}),
cannot be computed in conventional techniques that assumes casual
time evolution. To circumvent this difficulty we augment the standard
Keldysh technique by introducing two forward and two backward evolutions
shown in Fig. \ref{fig:The-augmented-Keldysh}b.

In order to describe the augmented technique we recall the conventional
Keldysh technique \cite{Keldysh1965} first. There, the differently
ordered correlators are given by 
\begin{equation}
N_{\alpha\beta\gamma\delta}(t)=\left\langle \mathcal{T}_{\mathcal{C}_{K}}\hat{O}_{\alpha}(t_{2})\hat{O}_{\beta}(t_{1})\hat{O}{}_{\gamma}(t_{2})\hat{O}{}_{\delta}(t_{1})\exp\left(-i\int_{\mathcal{C}_{K}}\hat{H}_{int}(t)dt\right)\right\rangle ,\label{eq:N_abcd}
\end{equation}
where $\alpha,\beta,\gamma,\delta=\pm$ denote the positions of the
operators on the traditional Keldysh contour shown in Fig. \ref{fig:The-augmented-Keldysh}a.
Here all observable operators $\hat{O}$ and the interaction part
of the Hamiltonian $\hat{H}_{int}$ are in the interaction representation,
the averaging is done with a density matrix (that represents the initial
conditions in the past), the symbol $\mathcal{T}_{C_{K}}$ denotes
ordering of the operators on the Keldysh contour, i.e. the operator
referring to the position down the contour is on the left in (\ref{eq:N_abcd})
(for fermion operators the change of order brings in minus sign).
One can see that by choosing indexes $\alpha,\beta,\gamma,\delta=\pm$
one can get different order of operators but never the anomalous order
required by Eq. (\ref{eq:A_general}).

The augmented contour $\mathcal{C}_{aK}$ allows anomalous order of
the operators such as in Eq. (\ref{eq:A_general}). For this contour
the indices $\alpha,\beta,\gamma,\delta$ can acquire four values,
$u\pm,d\pm$ (where $u$ stays for the up and $d$ for the down parts
of the contour). The expression similar to Eq. (\ref{eq:N_abcd})
\begin{equation}
A_{\alpha\beta\gamma\delta}(t)=\left\langle \mathcal{T}_{C_{aK}}\hat{O}_{\alpha}(t_{2})\hat{O}_{\beta}(t_{1})\hat{O}{}_{\gamma}(t_{2})\hat{O}{}_{\delta}(t_{1})\exp\left(-i\int_{\mathcal{C}_{aK}}\hat{H}_{int}(t)dt\right)\right\rangle ,\label{eq:A_abcd}
\end{equation}
with the choice $\alpha=u+$, $\beta=d+$,$\gamma=u-$, $\delta=d-$
becomes out-of-order correlator $\mathcal{A}(t)$. Clearly other combinations
of indices will produce normal as well abnormal correlators.

Equation (\ref{eq:A_abcd}) is the essence of the augmented technique.
Unitary evolution on $u/d$ segments of the contour can be viewed
as the evolution of the different worlds (we will use this term loosely
throughout the paper) with the same Hamiltonian and the same initial
conditions (``correlated worlds `` initially). The correlator (\ref{eq:A_abcd})
can be viewed as the response at time $t$ to the perturbation (source)
at time 0. When the sources are located at the same up/down parts
of the contour, the response is directly measurable, we shall refer
to these sources as 'physical'. All the other sources will be referred
to as 'unphysical'. Our ultimate objective is to describe how these
``correlated worlds'' become ``uncorrelated worlds'' provided
that a small local perturbation is seeded differently in the two worlds
(butterfly effect). \footnote{The two contour formalism of Ref. \cite{Ansari2016} is not suitable
for this purpose: in this formalism the worlds are uncorrelated from
the very beginning though the disordered potential acts the same on
the both worlds.}

In the next few sections we generalize the rules and the results of
Keldysh technique for the augmented Keldysh technique. We will see
that almost all rules are going through up to the kinetic equation
where the correlation function describing not only the occupation
numbers but also the measure of the correlation between the different
worlds.

\subsection{Augmented space and Green function}

Similarly to usual diagrammatic technique, we introduce the $4\times4$
matrix of Green functions of Fermi or Bose fields. It is convenient
to view this four dimensional space as a direct product of $2\times2$
Keldysh and $2\times2$ augmented space. Each operator (fermionic
or bosonic) $\psi(t)$ can be placed in four different points of contour
at time $t$, therefore it is enlarged into four dimensional vector:
\begin{equation}
\boldsymbol{\Psi}(1)=\left[\begin{array}{c}
\boldsymbol{\boldsymbol{\psi}}_{u}(1)\\
\boldsymbol{\boldsymbol{\psi}}_{d}(1)
\end{array}\right]_{a};\quad\boldsymbol{\psi}_{i}(1)=\begin{bmatrix}\psi_{i,+}(1)\\
\psi_{i,-}(1)
\end{bmatrix}_{K}\label{eq:psi-definition}
\end{equation}
where $(1),(2)$ are the short hand notations for the coordinates,
times (and might be spin) that specifies the single particle state:
$i\equiv(t_{i},r_{i},\sigma_{i})$. In these notations the $4\times4$
matrix Green function reads

\begin{equation}
i\hat{\mathcal{G}}(1,2)=\langle\mathcal{T_{\mathcal{C}}\boldsymbol{\Psi}}(1)\otimes\boldsymbol{\Psi}^{\dagger}(2)\rangle.\label{eq:G_definition}
\end{equation}
As usual, the components of the Green functions are linearly dependent.
This redundancy is eliminated by the Keldysh rotation, which is conveniently
described by the Pauli matrices 
\begin{align}
\hat{\tau}_{0}^{\cdot} & =\begin{pmatrix}1 & 0\\
0 & 1
\end{pmatrix};\ \hat{\tau}_{1}^{\cdot}=\begin{pmatrix}0 & 1\\
1 & 0
\end{pmatrix};\ \hat{\tau}_{2}^{\cdot}=\begin{pmatrix}0 & -i\\
i & 0
\end{pmatrix};\ \hat{\tau}_{3}^{\cdot}=\begin{pmatrix}1 & 0\\
0 & -1
\end{pmatrix};\nonumber \\
\hat{\tau}_{+}^{\cdot} & =\begin{pmatrix}0 & 1\\
0 & 0
\end{pmatrix};\ \hat{\tau}_{-}^{\cdot}=\begin{pmatrix}0 & 0\\
1 & 0
\end{pmatrix}.\label{eq:tau_matrices}
\end{align}
The superscript, $\cdot=a,K$, describes the space (augmented or Keldysh)
in which these matrices act. In terms of matrices (\ref{eq:tau_matrices})
the Keldysh rotation is given by \footnote{Such parametrization of the Keldysh space was first introduced by
Larkin and Ovchinnikov \cite{Larkin1975}.}\begin{subequations}

\begin{eqnarray}
\boldsymbol{\Psi} & \to & \hat{\mathcal{R}}\boldsymbol{\Psi,}\\
\boldsymbol{\Psi}^{\dagger} & \to & \boldsymbol{\bar{\Psi}=}\boldsymbol{\Psi}^{\dagger}\hat{\mathcal{R}}^{\dagger}\cdot\left(\hat{\tau}_{1}^{K}\otimes\hat{\tau}_{0}^{a}\right),\\
\hat{\mathcal{G}} & \to & \left(\hat{\mathcal{R}}\hat{\mathcal{G}}\hat{\mathcal{R}}^{\dagger}\right)\cdot\left(\hat{\tau}_{1}^{K}\otimes\hat{\tau}_{0}^{a}\right),\\
\hat{\mathcal{R}} & = & \exp\left(\frac{i\pi\hat{\tau}_{2}^{K}\otimes\hat{\tau}_{0}^{a}}{4}\right)
\end{eqnarray}

\label{eq:Keldysh_rotation}\end{subequations}

After rotation (\ref{eq:Keldysh_rotation}), the Green function acquires
the form 
\[
\hat{\mathcal{G}}=\left(\begin{array}{cc}
\hat{G}_{uu} & \hat{G}_{ud}\\
\hat{G}_{du} & \hat{G}_{dd}
\end{array}\right)_{a},
\]
where 
\begin{eqnarray}
\hat{G}_{uu} & = & \left(\begin{array}{cc}
G^{R} & G^{K}\\
0 & G^{A}
\end{array}\right)_{K},\,\hat{G}_{ud}=\left(\begin{array}{cc}
0 & \Gamma^{K}\\
0 & 0
\end{array}\right)_{K}\nonumber \\
\hat{G}_{du} & = & \left(\begin{array}{cc}
0 & \bar{\Gamma}^{K}\\
0 & 0
\end{array}\right)_{K},\thinspace\hat{G}_{dd}=\left(\begin{array}{cc}
\tilde{G}^{R} & \tilde{G}^{K}\\
0 & \tilde{G}^{A}
\end{array}\right)_{K}.\label{eq:After rotation}
\end{eqnarray}
In the absence of the non-physical sources\footnote{We distinguish physical sources that can be realized in experiment
and the non-physical ones that require time-machine for their implementation}, the components diagonal in the augmented space are equal, $\hat{G}_{uu}=\hat{G}_{dd}$,
and coincide with the conventional Green functions. In particular,
the retarded, advanced and Keldysh Green functions, $G^{R,A,K}$ are
given by 
\begin{eqnarray}
iG^{R}(\mbox{1,2\ensuremath{)}} & = & \left\langle \psi(1)\psi^{\dagger}(2)\pm\psi^{\dagger}(2)\psi(1)\right\rangle \theta(t_{1}-t_{2}),\nonumber \\
iG^{A}(\mbox{1,2\ensuremath{)}} & = & -\left\langle \psi(1)\psi^{\dagger}(2)\pm\psi^{\dagger}(2)\psi(1)\right\rangle \theta(t_{2}-t_{1}),\nonumber \\
iG^{K}(\mbox{1,2\ensuremath{)}} & = & \left\langle \psi(1)\psi^{\dagger}(2)\mp\psi^{\dagger}(2)\psi(1)\right\rangle ,\label{eq:GRGAGKstandard}
\end{eqnarray}
(hereinafter, the upper sign corresponds to fermions and lower to
bosons unless stated otherwise), whilst the inter-world functions
read: 
\begin{eqnarray}
\Gamma^{K}(1,2) & =-2i\left\langle \psi(1)\psi^{\dagger}(2)\right\rangle = & G^{K}(1,2)+[G^{R}(1,2)-G^{A}(1,2)],\label{eq:Gamma-equilibrium}\\
\bar{\Gamma}^{K}(1,2) & =\pm2i\left\langle \psi^{\dagger}(2)\psi(1)\right\rangle = & G^{K}(1,2)-[G^{R}(1,2)-G^{A}(1,2)].\nonumber 
\end{eqnarray}

In the absence of non-physical sources, these functions include the
information on the single particle spectrum and on the distribution
functions of holes (particles), $\Gamma^{K}$ ($\bar{\Gamma}^{K}$).
We note that even in the presence of non-physical sources the diagonal
components are not influenced by the non-diagonal ones. This is because
the correlations between the upper and down worlds can not affect
the dynamics in each of these worlds. Formally, this means that the
structure of the Green functions always retain the form of Eq.(\ref{eq:After rotation}).
Only the relation (\ref{eq:Gamma-equilibrium}) between diagonal and
non-diagonal components in augmented space can be modified by the
presence of non-physical sources. In fact, the violation of the relation
(\ref{eq:Gamma-equilibrium}) will be the formal indicator of the
quantum butterfly effect.

\subsection{Observables and computables\label{subsec:Observables-and-computables}}

We distinguish the correlators (\emph{observables}) that can be in
principle measured by a physics experiment and the ones that can only
be studied in the rather artificial system that allows inversion of
time directions. Because the latter can be more readily studied by
numerical simulations, where the unitary evolution can be formally
reversed\footnote{Butterfly effect in this case might appear due to the rounding errors
in back and forth evolutions.}, we call them \emph{computables}.

The example of observable is given by the casual correlator 
\begin{equation}
\mathcal{N}_{\rho\rho}(t)=\frac{1}{2}\left\langle T_{\mathcal{C}}\left(\boldsymbol{\bar{\Psi}}(t)(\tau_{1}^{K}\otimes\tau_{0}^{a})\boldsymbol{\Psi}(t)\right)\left(\boldsymbol{\bar{\Psi}}(0)(\tau_{0}^{K}\otimes\tau_{0}^{a})\boldsymbol{\Psi}(0)\right)\right\rangle \label{eq:N_rhorho}
\end{equation}
that describes the density response at time $t$ to the perturbation
at time $0$. Indeed the correlator (\ref{eq:N_rhorho}) rewritten
in terms of the original fields $\psi$ has the form 
\[
\mathcal{N}_{\rho\rho}(t)=\left\langle \left[\psi^{\dagger}(t,r)\psi(t,r),\psi^{\dagger}(0,r_{0})\psi(0,r_{0})\right]\right\rangle ,
\]
the usual rules of linear response imply that the density induced
by the scalar potential applies at point $r_{0}$ at time $t=0$ is
given by $-i\mathcal{N}_{\rho\rho}(t)$. In fact this structure is
general for Keldysh technique: the physical perturbation comes with
$\tau_{0}^{K}$ while the observable comes with $\tau_{1}^{K}$.

In contrast the out-of-time-ordered correlator provides the example
of the computable. In this paper we focus on out-of-time-ordered correlators
of the form 
\begin{equation}
\mathcal{A^{\gamma\delta}}_{\alpha\beta}(t,r;t'r')=\left\langle T_{\mathcal{C}}\left(\psi_{\alpha}(t,r)\psi_{\beta}^{\dagger}(t',r)\right)\left(\psi_{\gamma}^{\dagger}(0,0)\psi_{\delta}(0,0)\right)\right\rangle \label{eq:A_psi_general}
\end{equation}
that becomes out-of-time-ordered for many combinations of indices
$\alpha,$ $\beta,$ $\gamma$, $\delta$. For instance, for $\alpha=u+$,
$\beta=d+$,$\gamma=u-$, $\delta=d-$ it provides an example of the
general correlator (\ref{eq:A_abcd}) discussed in Section \ref{subsec:Augmented-Keldysh-technique}.
It is convenient to separate, as we have done here by parenthesis,
the 'source' term provided by the product of two operators at time
$t=0$ and the 'response term' provided by two operators at time $t\approx t'>0$.

For the fixed $\gamma$ and $\delta$ the correlator (\ref{eq:A_psi_general})
can be viewed as the Green function $G_{\alpha\beta}(t,r;t',r)$ computed
in the states modified by the action of the operators $\psi_{\gamma}(0,0)\psi_{\delta}^{\dagger}(0,0)$.
In particular, it satisfies the same identities as the Green function:
\begin{equation}
\mathcal{A}_{u+,d-}^{\gamma\delta}=\mathcal{A^{\gamma\delta}}_{u-,d-}=\mathcal{A}_{u-,d+}^{\gamma\delta}=\mathcal{A}_{u+,d+}^{\gamma\delta}.\label{eq:A_ident}
\end{equation}
After Keldysh rotation in indices $\alpha$ and $\beta$ the correlator
(\ref{eq:A_psi_general}) acquires the same general form as the Green
function (\ref{eq:After rotation}).

Because of the identities (\ref{eq:A_ident}) one can choose many
equivalent forms of the out-of-time-ordered correlators that display
unusual behavior. It will be more convenient to us to compute the
symmetrized correlator defined by 
\begin{equation}
\mathcal{A}_{\rho\rho}(t,t',r)=\left\langle T_{\mathcal{C}}\left(\boldsymbol{\bar{\Psi}}(t',r)(\tau_{1}^{K}\otimes\tau_{1}^{a})\boldsymbol{\Psi}(t,r)\right)\mathcal{\hat{S}}_{0}\right\rangle .\label{eq:A_rhorho}
\end{equation}

The source term, $\hat{\mathcal{S}}_{0}$, can have many equivalent
forms that distinguish upper and down Worlds, we can choose for instance
\begin{equation}
\hat{\mathcal{S}}_{0}=\psi_{u-}^{\dagger}(0,0)\psi_{d-}(0,0)\label{eq:S_0}
\end{equation}
This term destroys one particle in the down World and creates it back
before the evolution in the upper World starts (notice that for the
operators at $t=0$ $\psi_{u-}^{\dagger}(0)=\psi_{d+}^{\dagger}(0)$,
see Fig. \ref{fig:The-augmented-Keldysh}). As we shall see below
the final results depend very weakly on the particular form of the
source term.

The response operator in this correlator is the sum of four terms
\begin{equation}
\hat{\mathcal{R}}_{tt'}(r)=\boldsymbol{\bar{\Psi}}(t',r)(\tau_{1}^{K}\otimes\tau_{1}^{a})\boldsymbol{\Psi}(t',r)=\psi_{u-}^{\dagger}\psi_{d-}-\psi_{u-}\psi_{d-}^{\dagger}+\psi_{u+}^{\dagger}\psi_{d+}-\psi_{u+}\psi_{d+}^{\dagger}\label{eq:R_tt'}
\end{equation}
that measures the product of the distribution functions and the correlations
between the worlds. The minus sign in this equation is due to fermionic
commutation rules. 

As usual, any correlator allows for a pictorial representation to
facilitate the basic structure of the theory and to be able to sum
up the most important parts of the perturbative expansions up to infinite
order. We develop the diagrammatic rules for the technique in the
augmented space below in sections \ref{subsec:Basic-rules-of}, \ref{subsec:Vertices},
\ref{subsec:Diagram-technique:-summary}.

In the absence of the unphysical sources the two worlds remain perfectly
correlated. The stability of this solution can be discussed in very
general terms without the knowledge of the details of the microscopic
model. In fact, the existence of the self-energy, Dyson equation,
and the general thermodynamic relations are sufficient to prove that
the perfectly correlated solution is unstable. We begin with these
general considerations.

\subsection{Dyson equation.\label{subsec:Dysion equation.}}

In any field theory that allows the separation of the Hamiltonian
into bare ($H_{0})$ and interacting ($H_{int}$) parts, one can introduce
the notion of bare Green function, $\widehat{G}_{0}$, corresponding
to Hamiltonian $H_{0},$ the full Green functions (defined above)
and the self energies, $\hat{\Sigma}$, that take into account the
effects of the interaction on the bare Green functions. In diagramm
technique the self-energy can be defined as the sum of all one-particle-irreducible
diagrams (see Sections \ref{subsec:Basic-rules-of}, \ref{subsec:Vertices},
\ref{subsec:Diagram-technique:-summary}). The Green functions and
self-energies obey the Dyson equation that can be written in two equivalent
forms\begin{subequations} 
\begin{eqnarray}
(\hat{H}_{0}\tau_{0}^{a}-\hat{\Sigma})\circ\hat{G} & = & \hat{1},\\
\hat{G}\circ(\hat{H}_{0}\tau_{0}^{a}-\hat{\Sigma}) & = & \hat{1},
\end{eqnarray}
\label{eq:DysonRightLeft}\end{subequations} where $\hat{1}$ is
the unit operator in the space-time and the augmented Keldysh space,
the symbol $\circ$ implies the matrix multiplication in the augmented
space and the convolution in space-time. The operator $\hat{H}_{0}$
is diagonal in Keldysh and augmented spaces with the diagonal elements
defined by equations $\hat{H}_{0}G_{0}^{R}=1$, $\hat{H}_{0}G_{0}^{A}=1$,
it is related to the Hamiltonian $H_{0}$ by $\hat{H}_{0}=id/dt-H_{0}$
. 

The general structure of the Green functions (\ref{eq:After rotation})
implies that the parts of the Green function that are retarded and
advanced in Keldysh space remain diagonal in the augmented space.
Because the bare retarded and advanced Green functions are diagonal
in the augmented space, the self energies $\Sigma_{\alpha\beta}^{A,R}=\delta_{\alpha\beta}\Sigma_{\alpha}^{A,R}$
remain also diagonal and they are given by the solution of the equations
\begin{subequations}

\begin{eqnarray}
(\hat{H}_{0}-\Sigma_{\alpha}^{R/A})\circ G_{\alpha}^{R/A} & = & 1\\
G_{\alpha}^{R/A}\circ(\hat{H}_{0}-\Sigma_{\alpha}^{R/A}) & = & 1,\ \alpha=u,d.
\end{eqnarray}

\label{eq:GRA}\end{subequations}

As usual, the non-diagonal part of the Green functions in Keldysh
space is not entirely determined by the Eqs. (\ref{eq:GRA}): it also
depends on the initial conditions. Its evolution is described by the
homogeneous equations\begin{subequations}

\begin{eqnarray}
(\hat{H}_{0}-\Sigma_{\alpha}^{R})\circ G_{\alpha\beta}^{K}-\Sigma_{\alpha\beta}^{K}\circ G_{\beta}^{A} & = & 0,\\
G_{\alpha\beta}^{K}\circ(\hat{H}_{0}-\Sigma_{\beta}^{A})-G_{\alpha}^{R}\circ\Sigma_{\alpha\beta}^{K} & = & 0.
\end{eqnarray}
\label{eq:GK}\end{subequations}

Notice that both the diagonal and the non-diagonal parts $G_{\alpha\beta}^{K}$
are controlled by the initial conditions. We emphasize that the diagonal
components $\Sigma_{\alpha}^{A,R}$, $\Sigma_{\alpha\alpha}^{K}$
may depend on the diagonal components of the Green functions in the
augmented space, $G_{\alpha\alpha}^{K}$, but not on the other diagonal
(e.g. $G_{\beta}^{A,R}$, $G_{\beta\beta}^{K}$ $\beta\neq\alpha)$
or the non-diagonal ( $G_{\alpha\beta}^{K}$) Keldysh components.
This observation turns out to be the key of the description of the
instability in the evolution of non-diagonal correlations as we see
in the next subsection.

\subsection{Stability and instability.\label{subsec:Stability-and-instability.} }

Let us consider the fermionic Green function for the sake of concreteness.
The Keldysh components of the Green function can be conveniently parametrized
via 
\begin{equation}
G_{\alpha\beta}^{K}=G_{\alpha}^{R}\circ\mathcal{F}_{\alpha\beta}-\mathcal{F}_{\alpha\beta}\circ G_{\beta.}^{A}\label{eq:Fsdefinitions}
\end{equation}
For $\alpha=\beta$ this equation reduces to the conventional parametrization
of $G^{K}$ in terms of the quantum distribution function, $\mathcal{F}_{uu}$
and $\mathcal{F}_{dd}$; for $\alpha\neq\beta$ it gives the parametrization
of the new functions $\Gamma^{K}$ and $\bar{\Gamma}^{K}$ in terms
of $\mathcal{F}_{ud}$ and $\mathcal{F}_{du}$.

Substituting Eq.(\ref{eq:Fsdefinitions}) into Eqs. (\ref{eq:GK})
and using Eqs. (\ref{eq:GRA}), we find that Eqs. (\ref{eq:GK}) are
satisfied for $\mathcal{F}_{\alpha\beta}$ solving the quantum kinetic
equation 
\begin{equation}
H_{0}\circ\mathcal{F}_{\alpha\beta}-\mathcal{F}_{\alpha\beta}\circ H_{0}=\left[\Sigma_{\alpha}^{R}\mathcal{\circ F}_{\alpha\beta}-\mathcal{F}_{\alpha\beta}\circ\Sigma_{\alpha}^{A}\right]-\Sigma_{\alpha\beta}^{K}.\label{eq:QKineticEq}
\end{equation}
In the quasiclassical approximation the two terms in brackets correspond
to the outgoing scattering processes (this term taken alone always
leads to dissipation) whilst the last term corresponds to the incoming
processes (this term taken alone always leads to instability).

In thermal equilibrium the Green functions depend only on the time
difference. The diagonal parts of the electron self energies are related
by the fluctuation-dissipation theorem (FDT): 
\begin{equation}
\Sigma_{\alpha\alpha}^{K}(\epsilon)=\left[\Sigma_{\alpha}^{R}(\epsilon)-\Sigma_{\alpha}^{A}(\epsilon)\right]n_{0}(\epsilon),\quad n_{0}(\epsilon)=\tanh\left(\frac{\epsilon-\mu}{2T}\right)\label{eq:FDT}
\end{equation}
where $\epsilon$ is the frequency conjugated to the time difference,
$\mu$ is the chemical potential, and $T$ is the temperature: both
of them are determined by the initial conditions. For Bosons one should
replace $n_{0}(\epsilon)=\tanh(\ldots)$ by $p_{0}(\omega)=\coth(\dots)$.
For phonons (which number is not conserved) the chemical potential
$\mu=0$. In equilibrium the left hand side of Eq. (\ref{eq:QKineticEq})
is zero, substituting Eq. (\ref{eq:FDT}) into (\ref{eq:QKineticEq})
we see that $\left[1-\mathcal{F}_{uu}(\epsilon)\right]/2$ has the
meaning of the Fermi distribution function. The FDT also implies that
Eq. (\ref{eq:QKineticEq}) has a generally stable solution. The only
reason for this solution to become unstable is the metastability of
the state that might happen on the unstable branch of the phase transition.
However, even in this case, the ultimate fate of the system is a different
equilibrium characterized by different $\Sigma_{\alpha}^{R,A}$ and
\begin{equation}
\mathcal{F}_{uu}=\mathcal{F}_{dd}=n_{0}(\epsilon),\label{eq:F_uu}
\end{equation}
with different $\mu$ and $T$ that are found from the number of particle
and energy conservation in the new spectrum. That solution would be
stable again.

The stability of the thermal solution of Eq. (\ref{eq:QKineticEq})
is guaranteed by Boltzmann $H$-theorem and the global conservation
laws (energy and the number of particles). In the framework of Eq.
(\ref{eq:QKineticEq}) it means that for small deviations of $\mathcal{F}_{uu}$
, $\mathcal{F}_{dd}$ from the thermal distributions the outgoing
terms dominate incoming ones in Eq. (\ref{eq:QKineticEq}). This fact
is far from trivial because both terms are generically non-linear.

The equation (\ref{eq:QKineticEq}) allows for solution similar to
Eq. (\ref{eq:F_uu}) for the off-diagonal components: 
\begin{eqnarray}
\mathcal{F}_{ud} & = & 1+n_{0},\ \mathcal{F}_{du}=-1+n_{0}.\label{eq:F_ud}
\end{eqnarray}
We will call this solution the ``correlated worlds solution''.

In the Pauli matrix notation given by Eq. (\ref{eq:tau_matrices}),
Eq. (\ref{eq:F_uu}) and Eq. (\ref{eq:F_ud}) can be compactified
(for fermions) as 
\begin{equation}
\hat{\mathcal{F}}=n(\epsilon,p,r,t)\left(\hat{\tau}_{0}^{a}+\hat{\tau}_{1}^{a}\right)+i\hat{\tau}_{2}^{a},\label{eq:Fcorrelated-matrix}
\end{equation}
where anticipating further applications, we allow the distribution
function to depend not only on energy but also on the phase space
variable and time. The precise definition of the notion of the semiclassical
phase space is given in Sec. \ref{sec:Kinetic-equation-for}.

Notice that in contrast to the solution Eq. (\ref{eq:QKineticEq}),
the stability of the correlated worlds solution is not guaranteed
even if the solution Eq. (\ref{eq:F_uu}) is stable. Indeed, for the
diagonal (conventional) distribution function the small deviation
from equilibrium results in the non-zero RHS of Eq. (\ref{eq:QKineticEq})
in which outgoing terms always dominate incoming ones. Outgoing terms
always imply relaxation, which leads to the stability of the equilibrium
solution. For non-diagonal distribution function, the small deviation
from equilibrium results in the same incoming terms as for diagonal
distribution function but smaller outgoing terms. A small deviation
of diagonal term leads to two contributions to the outgoing term:
one due to the interaction induced change in $\Sigma_{\alpha}^{A,R}$,
another due to the change in $\mathcal{F}_{uu}$ ($\mathcal{F}_{dd}$)
itself. Because $\Sigma_{\alpha}^{A,R}$ do not depend on the off-diagonal
terms $\mathcal{F}_{ud}$ ($\mathcal{F}_{du}$), the former contribution
is missing for the off-diagonal terms. This makes deviation of the
outgoing term that tries to restore equilibrium smaller for non-diagonal
distribution function in the interacting system. Thus, for the off-diagonal
distribution function the outgoing terms do not necessarily dominate
the incoming ones for small deviations from the equilibrium. This
results in a possible instability of the solution of Eqs. (\ref{eq:F_ud}).
The computations for the specific models below show that this instability
is indeed present for electron-phonon and electron-electron interactions
but not for impurity scattering.

The alternative solution (allowed by conservation laws for off-diagonal
components) is 
\begin{equation}
\mathcal{F}_{ud}=\mathcal{F}_{ud}=0,\ {\textstyle or}\ \hat{\mathcal{F}}=n(\epsilon,p,r,t)\hat{\tau}_{0}^{a},\label{eq:zero}
\end{equation}
this solution will be called the ``uncorrelated worlds solution''.

The incoming term is second order (or higher) in $\mathcal{F}_{ud}$
($\mathcal{F}_{du}$), therefore it vanishes for the small deviation
from this solution. In contrast, the outgoing term is always linear
in $\mathcal{F}_{ud}$ ($\mathcal{F}_{du}$) and it dominates. Thus,
the uncorrelated worlds solution is generally stable and one expects
that the correlated worlds solution (\ref{eq:F_ud}) is not. The only
exception is the electron scattering by impurities that conserves
the number of particles at each energy separately. In this case both
outgoing and incoming terms are linear in all components of $\mathcal{F}_{\alpha\beta}$
and the previous arguments do not hold.

The meaning of the ``uncorrelated worlds solution'' (\ref{eq:zero})
is the following. Unlike their diagonal counterparts, $\mathcal{F}_{ud}$
($\mathcal{F}_{du}$) encode not only the distribution functions but
also the overlap of the many-body wave-functions evolving at the upper
and lower contour. Any decrease of this correlation diminishes the
values of both $\mathcal{F}_{ud}$ ($\mathcal{F}_{du}$). The proposed
instability is therefore nothing but the quantum butterfly effect,
the decay of $\mathcal{F}_{ud}$ ($\mathcal{F}_{du}$) everywhere
in the system results in the loss of the coherence between many body
wave functions describing upper and down Worlds. The ultimate solution
given in Eq. (\ref{eq:zero}) corresponds to the complete destruction
of the coherence between lower and upper contour.

The description of the evolution of the system from the ``correlated
worlds solution'' (\ref{eq:Fcorrelated-matrix}) to the ``uncorrelated
worlds solution'' (\ref{eq:zero}) is the subject of the further
sections.

\subsection{Basic rules of the diagram technique: Green functions\label{subsec:Basic-rules-of}}

\begin{figure}[h]
\includegraphics[width=1\textwidth]{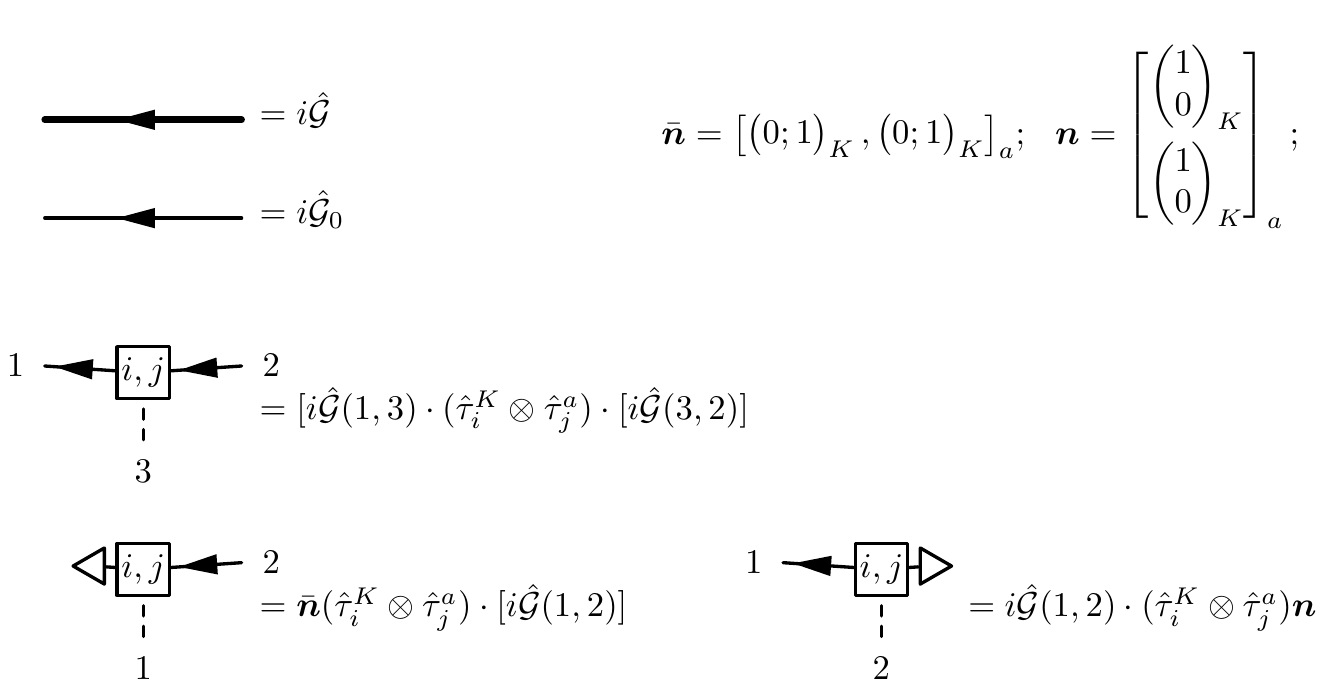}\centering

\caption{Definition of the basic elements of the diagrammatic technique. (a)
The lines (thick and thin) describe the exact and bare Green functions
respectively. (b) The box describes the matrix structure of the vertices.
(c) The vertices which do not conserve the number of particles (for
example absorption and emission of the phonons or photons). \label{fig:Definition-of-the}}
\end{figure}

\begin{figure}[t]
\includegraphics[width=0.5\textwidth]{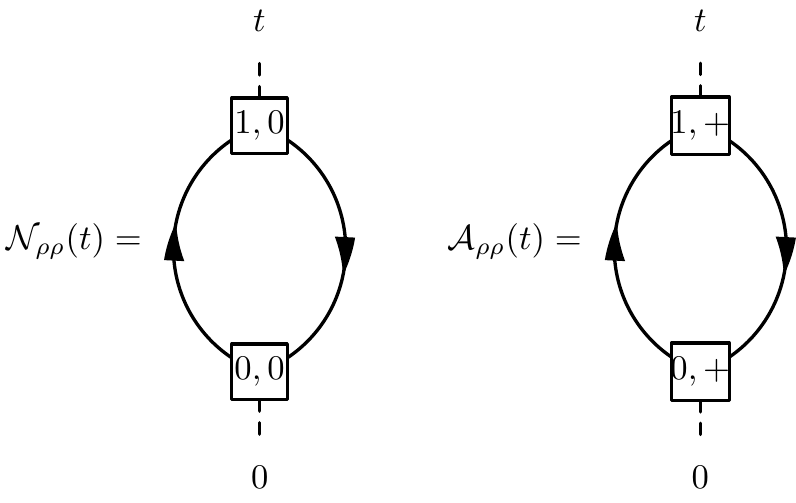}\centering

\caption{Diagrammatic expressions for the correlators (\ref{eq:N_rhorho})
and (\ref{eq:A_rhorho}) in non-interacting problem.\label{fig:Diagrammatic-expressions-for}}
\end{figure}

The basic elements of the diagrammatic representation needed to compute
the correlators in the augmented space are shown in Fig. \ref{fig:Definition-of-the}.
Notice that for keeping track of the Keldysh structure putting arrows
on the Green function for the real fields (as it is done throughout
this paper) is convenient but not necessary. In the absence of interactions
the observable (\ref{eq:N_rhorho}) and the computable (\ref{eq:A_rhorho})
are given by the diagrams shown in Fig. \ref{fig:Diagrammatic-expressions-for}.

Introduction of the separate notation for the box (see Fig. \ref{fig:Diagrammatic-expressions-for}
) enables one to display the matrix structure of the interaction vertices
as well.

\subsection{Vertices\label{subsec:Vertices}}

In order to develop the perturbation theory one needs to supplement
the expression for the Green functions with the expression for the
bare vertices. Because the unitary evolution in each sector is formally
independent, these vertices do not couple different sectors of the
augmented space. In the Keldysh space they have the usual structure.

\begin{figure}[h]
\includegraphics[width=0.8\textwidth]{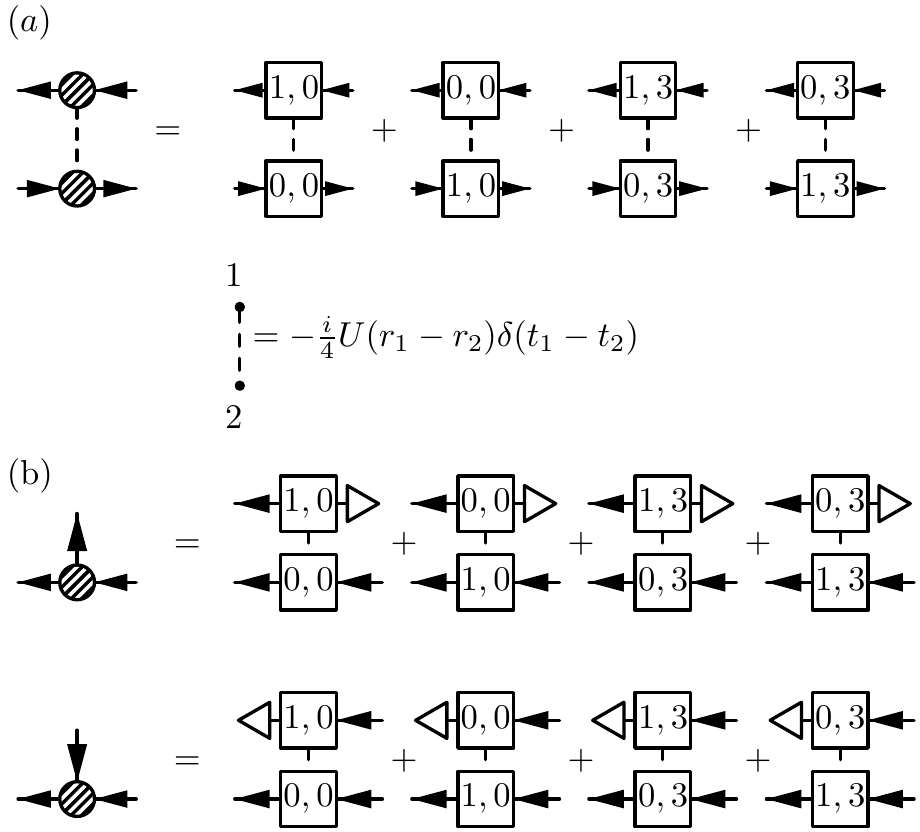}\centering

\caption{Vertex structure for (a) electron-electron interaction; and (b) for
the electron-phonon interaction. The outgoing or ingoing vertical
arrow in (b) can be any bosonic line including phonons, photons, impurity
potential etc. The short dashed lines on panel (b) are the delta function
in space and time.\label{fig:Vertex-structure-for-electron-electron}}
\end{figure}

To illustrate this point, more for the benefit of the readers familiar
with the conventional Keldysh technique, let us consider the textbook
\cite{RammerBook} example of the perturbation theory for the electrons
interacting with phonons. The lowest order contribution to the electron
self-energy has the form (formal general rules for the diagram techniques
will be summed in the next subsection): \begin{subequations} 
\begin{eqnarray}
\Sigma_{uu}^{ab} & = & i\lambda^{2}\gamma_{aa'}^{k}G_{uu}^{a'b'}D_{uu}^{kk'}\tilde{\gamma}_{b'b}^{k'},\\
\Sigma_{ud}^{ab} & = & i\lambda^{2}\gamma_{aa'}^{k}G_{ud}^{a'b'}D_{ud}^{kk'}\tilde{\gamma}_{b'b}^{k'},\\
\Sigma_{ud}^{ab} & = & i\lambda^{2}\gamma_{aa'}^{k}G_{ud}^{a'b'}D_{ud}^{kk'}\tilde{\gamma}_{b'b}^{k'},\\
\Sigma_{dd}^{ab} & = & i\lambda^{2}\gamma_{aa'}^{k}G_{dd}^{a'b'}D_{dd}^{kk'}\tilde{\gamma}_{b'b}^{k'},
\end{eqnarray}
\label{eq:Sigma-explicit}\end{subequations} where \begin{subequations}
\begin{eqnarray*}
\gamma_{ij}^{1} & = & \tilde{\gamma}_{ij}^{2}=\frac{1}{\sqrt{2}}\left[\hat{\tau_{0}}^{K}\right]_{ij},\\
\gamma_{ij}^{2} & = & \tilde{\gamma}_{ij}^{1}=\frac{1}{\sqrt{2}}\left[\hat{\tau_{1}}^{K}\right]_{ij}.
\end{eqnarray*}
\label{eq:gamma_ab}\end{subequations} The diagonal components in
the augmented space coincides with the ones for the regular technique.
The non-diagonal ones are found by using the Wick's theorem and noticing
that the vertices by themselves do not mix different sectors of the
augmented space. The matrix structure displayed by Eq. (\ref{eq:Sigma-explicit})
can be further compactified by equation 
\begin{equation}
\hat{\Sigma}_{\alpha\beta}=i\Upsilon_{\alpha\alpha'}^{\gamma}G_{\alpha'\beta'}\left(\frac{\lambda^{2}D_{\gamma\gamma'}}{4}\right)\tilde{\Upsilon}_{\beta'\beta}^{\gamma'},\label{eq:upsilon1}
\end{equation}
where $4\times4\times4$ matrices $\Upsilon_{\alpha\alpha'}^{\gamma}$
are given by \begin{subequations} 
\begin{eqnarray}
\Upsilon_{(i,j),(i',j')}^{(i'',j'')} & = & \sum_{\substack{l=0,1\\
m=0,3
}
}\left[\bar{n}\left(\tau_{l}^{K}\otimes\tau_{m}^{a}\right)\right]_{(i'',j'')}\left[\left(\tau_{\bar{l}}^{K}\otimes\tau_{m}^{a}\right)\right]_{(i,j),(i',j')},\\
\tilde{\Upsilon}_{(i,j),(i',j')}^{(i'',j'')} & = & \sum_{\substack{l=0,1\\
m=0,3
}
}\left[\left(\tau_{l}^{K}\otimes\tau_{m}^{a}\right)n\right]_{(i'',j'')}\left[\left(\tau_{\bar{l}}^{K}\otimes\tau_{m}^{a}\right)\right]_{(i,j),(i',j')}.
\end{eqnarray}
\label{eq:Upsilon_def}\end{subequations} Here $\bar{0}\equiv1$,
$\bar{1}=0$, $\bar{n}=(0,1;0,1)$, $n=(1,0;1,0)^{T}$, and the sum
over index $m=0,3$ gives the sum $\tau_{0}^{a}\otimes\tau_{0}^{a}+\tau_{3}^{a}\otimes\tau_{3}^{a}$
that is different from zero (and equal $2$) only for coinciding indices
in the augmented space. Pictorially, this matrix structure can be
summarized by Fig. \ref{fig:Vertex-structure-for-electron-electron}
where the basic blocks (boxes and triangles) are again defined in
Fig. \ref{fig:Definition-of-the}. The appearance of the vector $n$
in the formalism is due to the non-conservation of the number of particles
(phonons), see Fig. \ref{fig:Definition-of-the} c).

Note, that the same vertex $\Upsilon$ describes the interaction of
the electron with the disorder potential. The only difference is that
there is no time dependence of the quenched disorder potential, thus
the impurity line connecting different branches of the augmented Keldysh
contour never decays. The general structure of the vertices for electron-electron
interaction is shown in Fig. \ref{fig:Vertex-structure-for-electron-electron}a.
Note that is is described by the same building blocks as the electron-phonon
and electron-impurity interaction. Because the number of particles
in the electron-electron interaction is conserved the vector $n$
does not appear in this case.

\begin{figure}[t]
\includegraphics[width=0.8\textwidth]{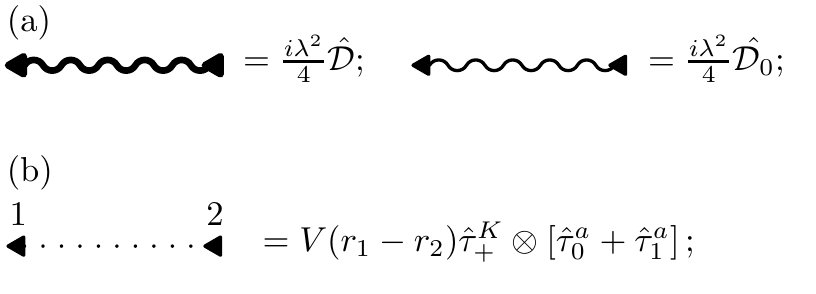}\centering

\caption{Remaining basic elements of the diagram technique for electron-phonon
a) and electron-impurity b) interactions. In (a) we include the interaction
constant $\lambda$into definition of the propagator to keep the vertices
of Fig. \ref{fig:Vertex-structure-for-electron-electron} intact.
Correlation function $V(r)$ describes the fluctuations due to the
weak random impurities. \label{fig:Basic-elements-of-the-diagram}}
\end{figure}

The definition of vertices has to be supplied with the remaining bosonic
Green functions, defined in Fig. \ref{fig:Basic-elements-of-the-diagram}

\subsection{Diagram technique: summary\label{subsec:Diagram-technique:-summary}}

We are now prepared to formulate the general rules of the diagram
technique that operates with the blocks defined in sections \ref{subsec:Basic-rules-of},
\ref{subsec:Vertices}:

In order to compute the correlator (observable or computable) one
has: (i) to place the sources and the interaction vertices, (ii) to
connect them by the Green function lines, (iii) to trace over the
indices in the augmented Keldysh space, (iv) to integrate over positions
of the interaction vertices, (v) to multiply the result by $(-1)^{N_{L}^{F}}$
, where $N_{L}^{F}$ is the number of the closed fermionic loops.

\begin{figure}
\includegraphics[width=0.8\textwidth]{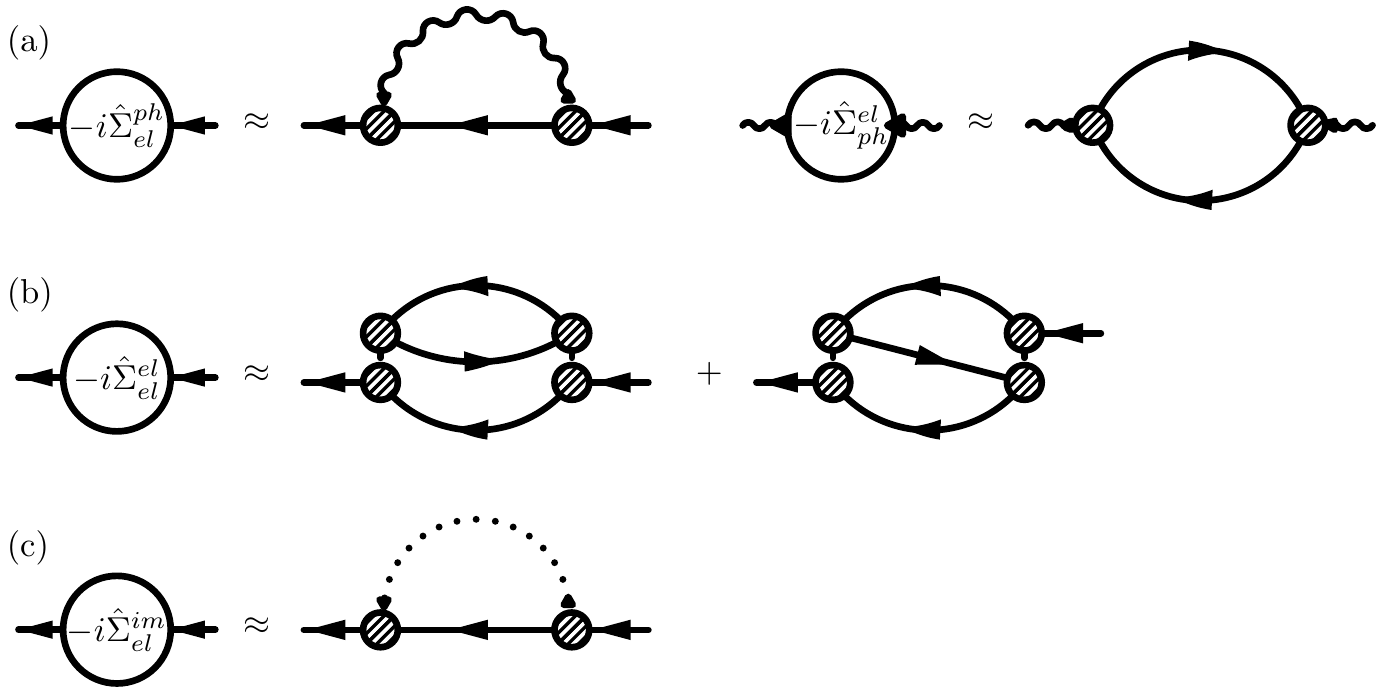}\centering

\caption{Self-energies for the (a) electron-phonon interaction, (b) electron-electron
interaction; and (c) electron in the Gaussian disordered potential.
The first order self-energy for the electron electron interaction
is discarded as not-related to the collisions but rather renormalizing
the self-consistent spectrum for the deterministic motion. The inside
lines for the Green function are solid which means that infinite series
of the rainbow diagrams is summed. This approximation is known to
lead to the quasi-classical Boltzmann equation and can be justified
for weak interactions or small enough disorder strength (neglecting
localization effects).\label{fig:Self-energies}}
\end{figure}

As usual, in order to derive the physical properties at large scales
one introduces the notion of self-energy that is defined as the sum
of all one-particle-irreducible diagrams. For example, the self-energies
for the electron-phonon, electron-electron and electron-disorder interaction
are shown in Fig. \ref{fig:Self-energies}.

\section{Microscopic Models\label{sec:Microscopic-Models}}

The instability expected in section \ref{subsec:Stability-and-instability.}
is of kinetic nature. Its form depends on the detailed form of the
kinetic equation and thus on the microscopic model on which the latter
is based. In the following we describe the models that allow one to
study the development of the instability in detail.

In all these models the main ingredient are mobile electrons that
form a Fermi sea. They are described by the quadratic Hamiltonian
\begin{equation}
H_{el}=\sum_{p}\xi_{p}\psi_{p}^{\dagger}\psi_{p}\label{eq:H_el}
\end{equation}
and characterized by the the bare Green function 
\begin{equation}
G_{0}^{R}=\frac{1}{\epsilon-\xi_{p}+i0},\ G_{0}^{A}=\frac{1}{\epsilon-\xi_{p}-i0},\label{eq:GRGA0}
\end{equation}
where the single particle energy $\xi_{p}$ is counted from the Fermi
energy $\epsilon_{F}$. The condition $\xi_{p}=0$ defines the Fermi
surface of the electrons.

For electrons the operator $H_{0}$ introduced in Eqs. (\ref{eq:DysonRightLeft})
acquires the form 
\begin{equation}
H_{0}=i\hbar\frac{\partial}{\partial t}-\xi_{p},\ p=-i\hbar\frac{\partial}{\partial r}.\label{eq:H_0}
\end{equation}
Here we restored the units of $\hbar$ for future convenience in developing
the quasiclassical approximation later on.

The three models for the electron interaction that we formulate below
differ by their conservation laws. The primitive model of electron-phonon
interaction (section \ref{subsec:Electron-phonon-interaction}) preserves
the total energy of the system and the number of electrons but not
the momentum of the system. The electrons in the impurity potential
(section \ref{subsec:Electrons-in-disorder}) is not a translational
invariant system, however, the scattering by impurities conserves
the energy of individual electrons, leading to infinite number of
conservation laws in this problem. The electron-electron interaction
(section \ref{subsec:Electron-electron-interaction}) preserves both
the translational and Galilean invariance, so it conserves the total
energy, momentum, and the particle number.\footnote{We neglect the effects of interference of different processes responsible
e.g. for the onset of the localization in disordered systems \cite{Altshuler1983}
or renormalization of the electron-phonon interaction strength by
disorder. The justification for this omission is that the effects
considered in this paper are dramatic already on the level of the
kinetic equation approximation. }

\subsection{Electron-phonon interaction\label{subsec:Electron-phonon-interaction}}

The simplest interacting model is the one in which the electrons interact
with dispersionless phonons with frequency $\omega_{0}$ (Einstein
phonons) with Hamiltonian 
\begin{equation}
H_{ph}=\sum_{r}\hbar\omega_{0}b_{r}^{\dagger}b_{r},\label{eq:phononHamiltonian}
\end{equation}
To avoid inconsequential consideration of the band structure we simply
assume (somewhat artificially) that all the points $r$ are random
and dilute, their density per unit volume is $n_{ph}.$ Notice that
$n_{ph}$ represents the density of phonon sites, the density of thermally
excited phonons is the product of $n_{ph}$ and phonon occupation
number. Bosons are interacting with electrons via 
\begin{equation}
H_{el-ph}=\sum_{r}u_{r}\psi^{\dagger}(r)\psi(r),\label{eq:electron-phonon-Hamiltonian}
\end{equation}
where $u_{r}=\lambda(b_{r}+b_{r}^{\dagger})$. Correlators of field
$u$ are given by the Green functions local in space, 
\begin{equation}
\hbar D_{0}^{R}=\frac{2\omega_{0}n_{ph}\delta(r_{1}-r_{2})}{(\omega+i0)^{2}-\omega_{0}^{2}},\ \hbar D_{0}^{A}=\frac{2\omega_{0}n_{ph}\delta(r_{1}-r_{2})}{(\omega-i0)^{2}-\omega_{0}^{2}}.\label{eq:DRDA}
\end{equation}
Poles at positive frequencies in these functions correspond to phonon
emission and at negative frequencies to phonon absorption (while the
physical energies of phonons are of course positive). These Green
functions should be used as the basic elements of the diagram technique
shown in Fig. \ref{fig:Basic-elements-of-the-diagram}a. Here and
below we adopt the traditional convention in which the phonon frequency
is denoted by $\omega$ whilst reserving $\epsilon$ for the electron
energy.

For phonons the operator $H_{0}$, introduced in Eq. \ref{eq:DysonRightLeft},
acquires the form 
\begin{equation}
H_{0}=-\frac{\partial^{2}}{\partial t^{2}}-\omega_{0}^{2}.\label{eq:phononsH0}
\end{equation}

Similarly to the electrons, see Eq. (\ref{eq:Fsdefinitions}), the
Keldysh part of the phonon Green function can be parametrized by 
\begin{equation}
D_{\alpha\beta}^{K}=D_{\alpha}^{R}\circ\mathcal{P}_{\alpha\beta}-\mathcal{P}_{\alpha\beta}\circ D_{\beta}^{A}\label{eq:Psdefinition}
\end{equation}
.

With this parametrization the form of the kinetic equation (\ref{eq:QKineticEq})
for the phonons remains the same;, the only difference is that their
equilibrium distribution functions for the correlated world solution
is 
\begin{equation}
\hat{\mathcal{\mathcal{P}}}=p_{ph}(\omega;r,t)\left(\hat{\tau}_{0}^{a}+\hat{\tau}_{1}^{a}\right)+i\hat{\tau}_{2}^{a},\label{eq:distributions-phonon-thermal}
\end{equation}
and for the uncorrelated world solution: 
\begin{equation}
\hat{\mathcal{\mathcal{P}}}=p_{ph}(\omega;r,t)\hat{\tau}_{0}^{a},\label{eq:p-uncorrlated}
\end{equation}
In the thermal equilibrium $p_{ph}=p_{0}(\omega).$

\subsection{Electrons in disorder potential\label{subsec:Electrons-in-disorder}}

The interaction with quenched disorder potential is described by the
Hamiltonian: 
\[
H_{el-imp}=\sum_{r}U(r)\psi_{r}^{\dagger}\psi_{r}
\]
After averaging over the disorder potential with correlator $\left\langle U(r)U(r')\right\rangle =V(r-r')$,
the translation invariance is restored for averaged correlation functions
and the diagrams for the electron correlators become similar to those
for electron-phonon interaction that carries zero frequency, see Fig.
\ref{fig:Basic-elements-of-the-diagram}.

\subsection{Electron-electron interaction\label{subsec:Electron-electron-interaction}}

The interaction between electrons is given by 
\begin{equation}
H_{el-el}=\frac{1}{2}\int\psi_{r}^{\dagger}\psi_{r'}^{\dagger}\psi_{r'}\psi_{r}U(r-r')drdr'\label{eq:H_el-el}
\end{equation}
The rules of the diagram technique are given in Fig. \ref{fig:Vertex-structure-for-electron-electron}.
In the discussion of the properties of this model we shall neglect
the spin of the electrons. For completeness, we also mention that
in the perturbation theory based on Eq. (\ref{eq:H_el-el}) the singular
terms proportional to $G^{K}(t,t)$ have to be understood as $G^{K}(1,1)\rightarrow2i\left\langle \Psi^{\dagger}(r_{1})\Psi(r_{1})\right\rangle $,
and $G^{R,A}(t,t)\to0$. Such terms appear only in the Hartree-Fock
contributions to the single electron spectrum and not in the collisions
interesting for us.

\section{Kinetic equation for normal and augmented Keldysh functions.\label{sec:Kinetic-equation-for}}

\subsection{Quasiclassical descriptions\label{subsec:Quasiclassical-descriptions}}

The kinetic equation (\ref{eq:QKineticEq}) fully determines the evolution
of the observables and the computables. However, it is not solvable
in a general case. Substantial simplification occurs in the quasiclassical
limit in which equations (\ref{eq:QKineticEq}) become local in time
and phase space. This simplification is possible if the rate of the
electron scattering is smaller than the relevant energy scales in
the problem: temperature for electron-electron or electron-phonon
interactions or Fermi energy for electrons in disorder potential.
For the diagonal part of the kinetic equation this is well established
and the theory of quantum corrections is well developed, see ref.
\cite{kamenev2011} for a pedestrian introduction. In the following
we shall assume that these conditions hold and that the quasiclassical
kinetic equation follows for the diagonal terms. Under these conditions
similar local equations hold for non diagonal Greem functions despite
the fact that these functions do not have classical meaning.

We follow the standard procedure for the derivation of the quasiclassical
equations for both diagonal and non-diagonal components. Any function
of two coordinates and two times can be represented as a Wigner transformation
\begin{equation}
W(t_{1,}r_{1};t_{2},r_{2})=\int w(\epsilon,p;t,r)e^{ipr_{-}/\hbar-i\epsilon t_{-}/\hbar}\frac{d\epsilon d^{d}p}{(2\pi\hbar)^{d+1}}\label{eq:Wigner_Transform}
\end{equation}
where $t=(t_{1}+t_{2})/2$, $r=(r_{1}+r_{2})/2$, $t_{-}=t_{1}-t_{2}$,
$r_{-}=r_{1}-r_{2}$. In this section we chose to keep the Planck
constant $\hbar$ explicitly so that the parameter for semiclassical
expansion is always displayed.

Using this representation for the Green functions of electrons and
employing Eq. (\ref{eq:H_0}) we get for the left hand side (LHS)
of the quantum kinetic equation (\ref{eq:QKineticEq}) 
\begin{equation}
H_{0}\circ\mathcal{F}_{\alpha\beta}-\mathcal{F}_{\alpha\beta}\circ H_{0}=i\hbar\left\{ \frac{\partial}{\partial t}+\frac{\partial}{\partial r}\frac{d\xi}{dp}\right\} \mathcal{F}_{\alpha\beta}(\epsilon,p;r,t)\label{eq:KinEqLHS}
\end{equation}
which coincides with the LHS of the classical Boltzmann equation for
both diagonal and off-diagonal components of the Green functions.

Similar arguments for the phonons lead to the LHS of the kinetic equation
\begin{equation}
H_{0}\circ\mathcal{P}_{\alpha\beta}-\mathcal{P}_{\alpha\beta}\circ H_{0}=i\hbar\omega\frac{\partial}{\partial t}\mathcal{P}_{\alpha\beta}(\omega,p;r,t).\label{eq:KinEqLTSph}
\end{equation}

The LHS of Eqs. (\ref{eq:KinEqLHS},\ref{eq:KinEqLTSph}) represent
the deterministic (Liouville) evolution corresponding to the unitary
quantum dynamics which is identical for both diagonal and non-diagonal
components in the augmented space. The right hand side (RHS) of the
kinetic equation describes the non-reversible probabilistic parts
and it is different for different models. The equations 
\begin{eqnarray}
\left\{ \frac{\partial}{\partial t}+\frac{\partial}{\partial r}\frac{d\xi}{dp}\right\} \mathcal{F}_{\alpha\beta}(\epsilon,p;r,t) & = & \mathrm{\left[St_{el}^{\cdots}\right]_{\alpha\beta}}\label{eq:KinEq_El_LHS}\\
\omega\frac{\partial}{\partial t}\mathcal{P}_{\alpha\beta}(\omega,p;r,t) & = & \omega\mathrm{\left[St_{ph}^{\cdots}\right]_{\alpha\beta}}\label{eq:KinEq_Ph_LHS}
\end{eqnarray}
describe the time evolution of the distribution functions. Here $\left[\mathrm{St}{}_{el}^{\cdots}\right],\,\left[\mathrm{St}{}_{ph}^{\cdots}\right]$
denote collision integrals for the particles (electrons or phonons)
scattered by other particles (denoted by $\cdots$). These collisions
integrals will be computed in the next section.

\subsection{Collision integrals for the specific models\label{subsec:Collision-integrals-for}}

The RHS of the kinetic equation allows a number of simplifications
in the leading order in $\hbar.$ Furthermore, below we shall consider
only the leading order term in the interation.

In the leading approximation one can replace 
\[
G^{R}-G^{A}=-2\pi i\delta(\epsilon-\xi_{p})
\]
and, with the same accuracy, 
\[
G_{\alpha\beta}^{K}=-2\pi i\delta(\epsilon-\xi_{p})\mathcal{F}_{\alpha\beta}.
\]

Analogously for phonons 
\begin{eqnarray*}
D^{R}-D^{A} & = & -2\pi in_{ph}\left[\delta(\omega-\omega_{0})-\delta(\omega+\omega_{0})\right]\\
D_{\alpha\beta}^{K} & = & -2\pi i\left[\delta(\omega-\omega_{0})\mathcal{P}_{\alpha\beta}(\omega_{0})-\delta(\omega+\omega_{0})\mathcal{P}_{\alpha\beta}(-\omega_{0})\right].
\end{eqnarray*}
Note that the fact that $D_{\alpha\alpha}^{K}(\omega)$ is an odd
function of frequency allows the simultaneous description of phonon
emission and absorption processes by a single $\mathcal{P_{\alpha\alpha}}(\omega>0)>0$
. Substituting these forms into the RHS of the kinetic equation we
obtain the collision integrals, 
\begin{equation}
\mathrm{St_{\alpha\beta}}=\frac{1}{\hbar}\Im\left\{ \left[\Sigma_{\alpha}^{R}\mathcal{\circ F}_{\alpha\beta}-\mathcal{F}_{\alpha\beta}\circ\Sigma_{\beta}^{A}\right]-\Sigma_{\alpha\beta}^{K}\right\} ,\label{eq:St_alphabeta}
\end{equation}
for different models. In Eq. (\ref{eq:St_alphabeta}) we kept only
the real part of the collision integral, we discuss various approsimation
involved in its derivation of in more detail in section \ref{subsec:Additional-remarks}

\subsubsection{Electron-phonon scattering. }

We calculate the lowest order diagrams shown in Fig. \ref{fig:Self-energies}.
For our model one neglect the correlations between phonons at different
space locations, i.e. the blobs for fermionic loop in self-energy
shown in Fig. \ref{fig:Self-energies} a) correspond to coinciding
points (with density $n_{ph}$). This implies that the electron self-energy
and collision integral contain an extra factor $n_{ph}$ with respect
to the phonon ones. In the diagonal sector we obtain (we do not write
down the spatial and time coordinates as the semiclassical collision
integrals are local in those variables) 
\begin{align}
\mathrm{\left[St_{el}^{ph}\right]}_{\alpha\alpha} & =n_{ph}\int\frac{d\omega dP_{1}M(P;P_{1},\omega)}{(2\pi)(2\pi\hbar)^{(d+1)}}\nonumber \\
\times & \left\{ -\left[\mathcal{L}_{el}^{ph}\right]_{\alpha}(P_{1},\omega)\mathcal{F}_{\alpha\alpha}(P)+\left[\mathcal{P}_{\alpha\alpha}(\omega)\mathcal{F}_{\alpha\alpha}(P_{1})+1\right]\right\} ;\nonumber \\
\mathrm{\left[St_{ph}^{el}\right]}_{\alpha\alpha} & =\frac{1}{2}\int\frac{dPdP_{1}M(P;P_{1},\omega)}{(2\pi\omega)((2\pi\hbar)^{(d+1)}}\label{eq:ph-diagonal}\\
\times & \left\{ -\left[\mathcal{L}_{el}^{ph}\right]_{\alpha}(P_{1},\omega)\mathcal{P}_{\alpha\alpha}(\omega)+\left[1-\mathcal{F}_{\alpha\alpha}(P)\mathcal{F}_{\alpha\alpha}(P_{1})\right]\right\} ,\nonumber 
\end{align}
where we introduced functions

\begin{align}
\left[\mathcal{L}_{el}^{ph}\right]_{\alpha}(P_{1},\omega) & =\mathcal{P}_{\alpha\alpha}(\omega)+\mathcal{F}_{\alpha\alpha}(P_{1}),\nonumber \\
\left[\mathcal{L}_{ph}^{el}\right]_{\alpha}(P_{1},\omega) & =\mathcal{F}_{\alpha\alpha}(P)-\mathcal{F}_{\alpha\alpha}(P_{1}),\label{eq:L_ph}
\end{align}
that determine outgoing rate. This notation is useful as the same
quantity enters the equations for the non-diagonal part.

For off-diagonal ($\alpha\neq\beta$) we obtain

\begin{align}
\mathrm{\left[St_{el}^{ph}\right]}_{\alpha\beta} & =n_{ph}\int\frac{dP_{1}dQ_{1}M(P;P_{1},\omega)}{(2\pi)(2\pi\hbar)^{(d+1)}}\left\{ -\mathcal{L}_{el}^{ph}(P_{1},\omega)\mathcal{F}_{\alpha\beta}(P)+\mathcal{P}_{\alpha\beta}(\omega)\mathcal{F}_{\alpha\beta}(P_{1})\right\} ,\nonumber \\
\left[St_{ph}^{el}\right]_{\alpha\beta} & =\frac{1}{2}\int\frac{dPdP_{1}M(Q;P_{1},\omega)}{(2\pi\omega)(2\pi\hbar)^{(d+1)}}\left\{ -\mathcal{L}_{ph}^{el}(P_{1},\omega)\mathcal{P}_{\alpha\beta}(\omega)-\mathcal{F}_{\alpha\beta}(P)\mathcal{F}_{\beta\alpha}(P_{1})\right\} ,\label{eq:ph-nondiagonal}
\end{align}
where we introduced the short hand notation

\begin{equation}
2\mathcal{L}_{\dots}^{\dots}(P_{1},Q_{1})\equiv\left[\mathcal{L}_{\dots}^{\dots}\right]_{u}(P_{1},Q_{1})+\left[\mathcal{L}_{\dots}^{\dots}\right]_{d}(P_{1},Q_{1}).
\end{equation}
The form factors $M$ include the matrix elements, conservation laws
for the electron and phonons colliding with each other, and their
spectrum: 
\[
M=\frac{\lambda^{2}}{2\hbar}\left[\left(2\pi\hbar\right)^{d+1}\delta(\epsilon-\epsilon_{1}-\hbar\omega)\right]\left[2\pi\hbar\delta(\epsilon_{1}-\xi(p_{1}))\right]\left[2\pi\hbar\sum_{\pm}\pm\delta(\omega\mp\omega_{0})\right]
\]
where numerical factor $1/2$ includes the difference of $\mathcal{F},\mathcal{P}$
from the physical distribution function by a factor of two. We find
it is more convenient to keep $\epsilon,\omega$ as independent energies
connected by $\delta$-functions in $M$ with physical spectrum to
have the symmetric form for the conservation laws and use the $d+1$
dimensional momentum vector $P=\left(\epsilon,p\right)$.

Here comes an important observation. Even though Eq. (\ref{eq:ph-diagonal})
and Eq. (\ref{eq:ph-nondiagonal}) look similar, their properties
are very different. Indeed the diagonal part satisfies the electron
number conservation law

\begin{equation}
\int dP\left\{ \mathrm{\left[St_{el}^{ph}\right]}_{\alpha\alpha}\delta(\epsilon-\xi(p))\right\} =0,\label{eq:phonon-particles}
\end{equation}
and the total energy conservation.\footnote{Note that the momentum is not conserved because of the locality of
the phonon correlations even though the averaged system is formally
translationally invariant. } 
\begin{equation}
\int dP\left\{ \mathrm{\left[St_{el}^{ph}\right]}_{\alpha\alpha}\delta(\epsilon-\xi(p))\epsilon\right\} +n_{ph}\int d\omega\left\{ \mathrm{\left[St_{ph}^{el}\right]}_{\alpha\alpha}\hbar\omega\sum_{\pm}\pm\delta(\omega\mp\omega_{0})\right\} =0.\label{eq:phonon-momentum}
\end{equation}
Moreover, one can explicitly check that the time derivative of the
entropy 
\begin{eqnarray}
\frac{dS}{dt}=\int dP\left\{ \mathrm{\left[St_{el}^{ph}\right]}_{\alpha\alpha}\delta[\epsilon-\xi(p)]\ \ln\frac{1+\mathcal{F}_{\alpha\alpha}(P)}{1-\mathcal{F}_{\alpha\alpha}(P)}\right\}  & +\nonumber \\
+n_{ph}\int d\omega\left\{ \mathrm{\left[St_{ph}^{el}\right]}_{\alpha\alpha}\sum_{\pm}\pm\delta(\omega\mp\omega_{0})\ \ln\frac{1+\mathcal{P}_{\alpha\alpha}(\omega)}{\mathcal{P}_{\alpha\alpha}(\omega)-1}\right\}  & \geq & 0,\label{eq:phonons-Htheorem}
\end{eqnarray}
which is the microscopic manifestation of Boltzmann H-theorem. Equality
is reached only for thermal distribution functions for which it reduces
to Eq. (\ref{eq:phonon-particles}) and Eq. (\ref{eq:phonon-momentum})
.

These equations allow to prove that the only stable solution of the
kinetic equation is given by the thermal distribution functions and
all deviations from it decay (generally, exponentially). In contrast,
the non-diagonal part does not satisfy any of these properties or
conservation laws. As we already mentioned, this absence of conservation
laws and H-theorem will be the key to understand the instability of
the thermal non-diagonal distributions for correlated worlds (\ref{eq:Fcorrelated-matrix})
and (\ref{eq:distributions-phonon-thermal}) and their subsequent
evolution to non-correlated worlds (\ref{eq:zero}), (\ref{eq:p-uncorrlated}).
The discussion of the instability will be done in Secs. \ref{sec:Instability-of-the}
and \ref{sec:Spatial-structure-of}. In the remainder of this section,
we list the properties of the collision integrals for the other physical
models of Sec. \ref{sec:Microscopic-Models}

\subsubsection{Electron-electron scattering. \label{subsec:Electron-electron-scattering.}}

We calculate the lowest order diagram shown on Fig.\ref{fig:Self-energies}.
In the diagonal sector we obtain 
\begin{align}
\mathrm{\left[St_{el}^{el}\right]}_{\alpha\alpha} & =\int\frac{dP_{1}dP_{2}dP_{3}M(P,P_{1};P{}_{2}P_{3})}{(2\pi\hbar)^{3(d+1)}}\left\{ -\left[\mathcal{L}_{el}^{el}\right]_{\alpha}(P_{1},P_{2},P_{3})\mathcal{F}_{\alpha\alpha}(P)\right.\nonumber \\
+ & \left.\left[\mathcal{F}_{\alpha\alpha}(P_{3})+\mathcal{F}_{\alpha\alpha}(P_{2})-\mathcal{F}_{\alpha\alpha}(P_{1})-\mathcal{F}_{\alpha\alpha}(P_{1})\mathcal{F}_{\alpha\alpha}(P_{2})\mathcal{F}_{\alpha\alpha}(P_{3})\right]\right\} ,\label{eq:electron-diagonal}
\end{align}
where we denoted

\begin{equation}
\left[\mathcal{L}_{el}^{el}\right]_{\alpha}=\mathcal{F}_{\alpha\alpha}(P_{2})\mathcal{F}_{\alpha\alpha}(P_{3})-\mathcal{F}_{\alpha\alpha}(P_{1})\mathcal{F}_{\alpha\alpha}(P_{3})-\mathcal{F}_{\alpha\alpha}(P_{1})\mathcal{F}_{\alpha\alpha}(P_{2})+1.\label{eq:diagonal-L}
\end{equation}
As before, the form factors $M$ include the matrix elements, the
conservation laws for the electron and phonons colliding with each
other, and their spectrum: 
\[
M=\frac{\left|U_{p_{2}-p}-U_{p_{3}-p}\right|^{2}}{8\hbar}\left[\left(2\pi\hbar\right)^{d+1}\delta(P+P_{1}-P_{2}-P_{3})\right]\prod_{i=1}^{3}\left[2\pi\hbar\delta(\epsilon_{i}-\xi(p_{i}))\right]
\]
where numerical factor $1/8$ includes the difference of $\mathcal{F},$
from the physical distribution function by a factor of two, and exchange
symmetry of the final state. We find it is more convenient to keep
$\epsilon,\omega$ as independent energies connected by $\delta$-functions
in $M$ with physical spectrum to have the symmetric form for the
conservation laws and use the $d+1$ dimensional momentum vectors
$P=\left(\epsilon,p\right)$.

The collision integral (\ref{eq:electron-diagonal}) satisfies the
conditions similar to those for electron-phonon scattering 
\begin{equation}
\int dP\left\{ \mathrm{\left[St_{el}^{el}\right]}_{\alpha\alpha}\delta(\epsilon-\xi(p))\right\} =0,\label{eq:electron-particle}
\end{equation}
(particle conservation) 
\begin{equation}
\int dP\left\{ \mathrm{\left[St_{el}^{el}\right]}_{\alpha\alpha}\delta(\epsilon-\xi(p))P\right\} =0,\label{eq:electron-momentum}
\end{equation}
(electron number conservation) 
\begin{eqnarray}
\int dP\left\{ \mathrm{\left[St_{el}^{ph}\right]}_{\alpha\alpha}\delta(\epsilon-\xi(p))\ln\frac{1+\mathcal{F}_{\alpha\alpha}(P)}{1-\mathcal{F}_{\alpha\alpha}(P)}\right\}  & \geq & 0,\label{eq:el_HTheorem}
\end{eqnarray}
(entropy growth).

For off-diagonal ($\alpha\neq\beta$) we obtain

\begin{multline}
\mathrm{\left[St_{el}^{el}\right]}_{\alpha\beta}=\int\frac{dP_{1}dP_{2}dP_{3}M(P,P_{1};P{}_{2}P_{3})}{(2\pi\hbar)^{3(d+1)}}\times\\
\left\{ -\mathcal{L}_{el}^{el}(P_{1},P_{2},P_{3})\mathcal{F}(P)_{\alpha\beta}+\mathcal{F}_{\alpha\beta}(P_{2})\mathcal{F}_{\alpha\beta}(P_{3})\mathcal{F}_{\beta\alpha}(P_{1})\right\} ,\label{eq:el-el_nondiagonal}
\end{multline}
where once again 
\[
2\mathcal{L}_{el}^{el}(P_{1},P_{2},P_{3})\equiv\left[\mathcal{L}_{el}^{el}\right]_{u}(P_{1},P_{2},P_{3})+\left[\mathcal{L}_{el}^{el}\right]_{d}(P_{1},P_{2},P_{3}).
\]

Similarly to the electron-phonon interaction, the collision integral
(\ref{eq:el-el_nondiagonal}) for the non-diagonal term is non-linear
due to the incoming term. This leads to the instability of the thermal
non-diagonal distribution.

\subsection{Electron-impurity scattering\label{subsec:Electron-impurity-scattering}}

The collision integral for the electron-impurity scattering is linear
in the distribution function 
\begin{equation}
\mathrm{\left[St_{el}^{im}\right]}_{\alpha\beta}=\int\frac{dp_{1}M(p,p_{1})}{(2\pi\hbar)^{d}}\left\{ -\mathcal{F}_{\alpha\beta}(p)+\mathcal{F}_{\alpha\beta}(p_{1})\right\} ,\label{eq:el-im_collision}
\end{equation}
where 
\[
M(p,p_{1})=\frac{2\pi}{\hbar}\left|V_{p-p_{1}}\right|^{2}\delta(\xi_{p}-\xi_{p_{1}}).
\]
This implies that in the case of the impurity scattering the non-diagonal
components of the Keldysh function have the same time evolution as
the diagonal ones, so the solution in which it is equal to the thermal
equilibrium distribution is stable. Note that electrons in the impurity
potential is a chaotic system. In this respect it is not different
from the electron-phonon and the electron-electron interaction. Nevertheless,
the non-diagonal components are stable, in contrast to the models
with electron-phonon and electron-electron interactions. This results
in a very different behavior of the out-of-time-ordered correlators
in this system.

\subsection{Additional remarks\label{subsec:Additional-remarks}}

It is worthwhile to emphasize that the basic form of the kinetic equation
and the forthcoming conclusions are not limited to the lowest order
self-energy calculation. In particular, taking into account the commutators
of the self-energy with $\mathcal{F}_{\alpha\beta}$ results in well
controllable corrections to the LHS of the kinetic equation and has
the meaning of the self-consistent spectrum. The higher order expansion
improves the accuracy of the matrix elements in the collision integrals
and also produces the real processes involving larger number of particles.
Neither of those complications seem to affect the basic relations
of the diagonal and non-diagonal evolutions and we will not be dwelling
on them in this paper. The imaginary part of the non-diagonal elements
of collision integral neglected in Eq. (\ref{eq:St_alphabeta}) formally
appears due to the difference between the distribution functions in
upper and down Worlds:
\[
\Im\mathrm{St_{\alpha\beta}}=\frac{1}{\hbar}\Re\left[\Sigma_{\alpha}^{R}-\Sigma_{\beta}^{R}\right]\mathcal{F}_{\alpha\beta}
\]
This effect also disappears in the leading quasiclassical approximation
and does not affect the instability discussed in the next sections.
For instance, for electron-phonon model this term becomes
\[
\Im\mathrm{St_{\alpha\beta}}\propto\int d\xi\left[\Re D^{R}(\epsilon-\xi)\left(\mathcal{F}_{\alpha}(\xi)-\mathcal{F_{\beta}}(\xi)\right)+\Re G^{R}(\xi)\left(\mathcal{P}_{\alpha}(\xi)-\mathcal{P_{\beta}}(\xi)\right)\right]
\]
It disappears for the two Worlds in equilibrium. Furthermore, it is
zero if one World has extra particle density that resulted in the
spatially non-unform chemical potential. 

\section{Instability of the augmented Keldysh functions in zero dimensional
case\label{sec:Instability-of-the}}

In this section we study the instability in the systems in which the
spatial dependence of the correlation functions can be neglected.

\subsection{Instability in electron-phonon model.\label{subsec:Instability-in-electron-phonon} }

The Einstein phonon distribution function is characterized by just
two numbers in each sectors of the augmented space that are the values
of $\mathcal{P}_{\alpha\beta}(\pm\omega_{0})$. In thermal equilibrium
the two diagonal components are given by Eq. (\ref{eq:distributions-phonon-thermal}).
Because the instability in the non-diagonal sector does not affect
the diagonal one (Sec. \ref{subsec:Dysion equation.}), for simplification
we assume that the diagonal sector for both electrons and phonons
is in equilibrium. This assumption is not essential, and the thermal
function can be replaced to its non-equilibrium value without any
technical complications. Because the phonon field is real, the non-diagonal
sectors are related to each other by the symmetry 
\[
\mathcal{P}_{ud}(\omega_{0})=-\mathcal{P}_{du}(-\omega_{0}),
\]
and the state of the phonons is described by two parameters 
\begin{eqnarray*}
\mathcal{P}_{ud}(\omega_{0}) & = & \theta,\\
\mathcal{P}_{du}(\omega_{0}) & = & \bar{\theta.}
\end{eqnarray*}
The phonon scattering process does not depend on the electron momentum,
so we need to keep only the energy, $\epsilon=\xi(p)$, dependence
of the electron distribution function: 
\begin{eqnarray*}
\mathcal{F}_{ud}(\epsilon,p;t) & = & f(\epsilon,t),\\
\mathcal{F}_{du}(\epsilon,p;t) & = & -\bar{f}(\epsilon,t).
\end{eqnarray*}
.

Inserting these definitions in the kinetic equations (\ref{eq:KinEq_El_LHS},\ref{eq:KinEq_Ph_LHS},\ref{eq:ph-nondiagonal})
and performing integrals over momentum we obtain\begin{subequations}
\begin{align}
\tau\frac{\partial f}{\partial t} & =-L\left(\frac{\epsilon}{2T},\frac{\omega_{0}}{2T}\right)f+\left[\theta f(\epsilon-\omega_{0})+\bar{\theta}f(\epsilon+\omega_{0})\right],\\
\tau\frac{\partial\bar{f}}{\partial t} & =-L\left(\frac{\epsilon}{2T},\frac{\omega_{0}}{2T}\right)\bar{f}+\left[\bar{\theta}\bar{f}(\epsilon-\omega_{0})+\theta\bar{f}(\epsilon+\omega_{0})\right],\\
\eta\tau\frac{\partial\theta}{\partial t} & =-\theta+I_{-},\\
\eta\tau\frac{\partial\bar{\theta}}{\partial t} & =-\bar{\theta}+I_{+},
\end{align}
\label{eq:df/dt_el-ph}\end{subequations}where we introduced the
positive quantities 
\begin{align}
I_{\pm} & =\frac{1}{2\omega_{0}}\int d\epsilon f(\epsilon)\bar{f}(\epsilon\pm\omega_{0});\label{eq:I_pm}\\
L(x,y) & =2\coth\left(y\right)-\tanh\left(x+y\right)+\tanh\left(x-y\right).\label{eq:LI}
\end{align}

We also replaced $\hbar\omega_{0}\to\omega_{0}$ as the semiclassical
expansion is already completed. In deriving Eqs. (\ref{eq:df/dt_el-ph}),
we neglected the energy dependence of the electron density of states,
$\nu$, we denoted 
\begin{equation}
\frac{1}{\tau}=\frac{2\pi\nu n_{ph}\lambda^{2}}{\hbar},\label{eq:etatau}
\end{equation}
and introduced the dimensionless parameter 
\begin{equation}
\eta=\frac{n_{ph}}{\hbar\nu\omega_{o}}.\label{eq:eta}
\end{equation}

Eqs. (\ref{eq:df/dt_el-ph}) are further simplified in the limits
$\eta\gg1$ and $\eta\ll1$. In the former limit the phonon relaxation
is slow compared to electrons, in the latter the electron relaxation
is slower. The limit $\eta\ll1$ seems to be the most relevant for
physical situations (e.g. to describe electrons interacting with low
density TLS) and moreover it will enable to develop intuition for
analyzing the more involved kinetics of electron-electron collisions.
Thus we focus on this limit only. Because the relaxation of $\theta$
is much faster than that of $f$, we can solve the Eqs (\ref{eq:df/dt_el-ph}c,d)
for $\theta,$$\bar{\theta}$ in the stationary limit. We obtain\begin{subequations}
\begin{align}
\tau\frac{\partial f}{\partial t} & =-L\left(\frac{\epsilon}{2T},\frac{\omega_{0}}{2T}\right)f+\left[I_{-}f(\epsilon-\omega_{0})+I_{+}f(\epsilon+\omega_{0})\right],\\
\tau\frac{\partial\bar{f}}{\partial t} & =-L\left(\frac{\epsilon}{2T},\frac{\omega_{0}}{2T}\right)\bar{f}+\left[I_{+}\bar{f}(\epsilon-\omega_{0})+I_{-}\bar{f}(\epsilon+\omega_{0})\right].
\end{align}
\label{eq:phonon-instability-main}\end{subequations}

Eqs. (\ref{eq:phonon-instability-main}) and Eqs. (\ref{eq:I_pm},\ref{eq:LI})
form the complete set of equations describing the time evolution of
the non-diagonal components of the distribution function. However,
they are still non-linear and nonlocal in energy space.

The further analysis is separated into two regimes: ``classical''
$\omega_{0}\ll T$ and ``quantum'' $\omega_{0}\gg T$. The difference
between these regimes is expected on the physical grounds: in the
``classical'' regime a large number of excitations is already present,
therefore, one expects (and we will see that that it is indeed the
case) that the perturbation results in the evolution that leads to
the uncorrelated fixed point, $f=0$, $\bar{f}=0$ with the characteristic
time of the order of $\tau.$ In the quantum regime, one expects that
the characteristic time is determined by an exponentially small number
of excitations and it becomes infinite at zero temperature.

Generally, one expects that in the gapful systems at $T=0$ a small
perturbations cannot lead to any instability, in particular, these
systems cannot be chaotic, so that the scrambling time is infinite.
The exponential growth of the characteristic time at low temperatures
in the gapless system found here implies a smooth crossover between
the properties of the gapful and gapless systems at $T=0$ .

\subsubsection{Classical limit ($\omega_{0}\ll T$)\label{subsec:Classical-limit-()}}

At high temperatures we can neglect $\omega_{0}$ in $I_{\pm}$ (\ref{eq:I_pm})
and in the arguments of $f$ in Eq. (\ref{eq:phonon-instability-main}),
we can also approximate $L=2/y$ in Eq. (\ref{eq:LI}). We see then
that the form of $f(\epsilon)$ and $\bar{f(}\epsilon)$ dependencies
is not changed by the evolution. Therefore, we can look for the solution
of Eq. (\ref{eq:phonon-instability-main}) in the form \begin{subequations}
\begin{eqnarray}
f(\epsilon) & = & \phi(t)\left[1+\tanh(\epsilon/2T)\right],\\
\bar{f}(\epsilon) & = & \phi(t)\left[1-\tanh(\epsilon/2T)\right].
\end{eqnarray}
\label{eq:f(epsilon)}\end{subequations}We obtain that the function
$\phi(t)$ obeys the first order differential equation 
\begin{equation}
\tau\frac{\partial\phi}{\partial t}=-\frac{4T}{\omega_{0}}\left(\phi-\phi^{3}\right)\label{eq:dphi/dt_0D}
\end{equation}
that has the unstable fixed point at $\phi=1$ and the stable fixed
point at $\phi=0$. The solution of this equation takes the form

\begin{equation}
\phi(t)=\left(\frac{1}{1+\exp\left[(t-t_{d})/t_{*}^{cl}\right]}\right)^{1/2},\label{eq:phi(t)}
\end{equation}
where $t_{*}^{cl}=\tau\omega_{0}/8T$. This time dependence is typical
of the dissipative instabilities. The delay time, $t_{d},$ depends
only logarithmically on the initial conditions: $t_{d}=t_{*}^{cl}\left|\ln\left[1-\phi(0)\right]\right|$
whilst the decay time $t_{*}^{cl}$ coincides with the classical scattering
time of the electrons that is inversely proportional to the density
of phonon sites, $n_{ph}$, and the phonon occupation number, $T/\omega_{0}$.
Note that this classical time, $t_{*}^{cl}$ , is less than the energy
relaxation time, in the limit $\omega_{0}\ll T$.

\subsubsection{Quantum limit ($\omega_{0}\gg T$)\label{subsec:Quantum-limit-()}}

\begin{figure}
\includegraphics[width=0.8\textwidth]{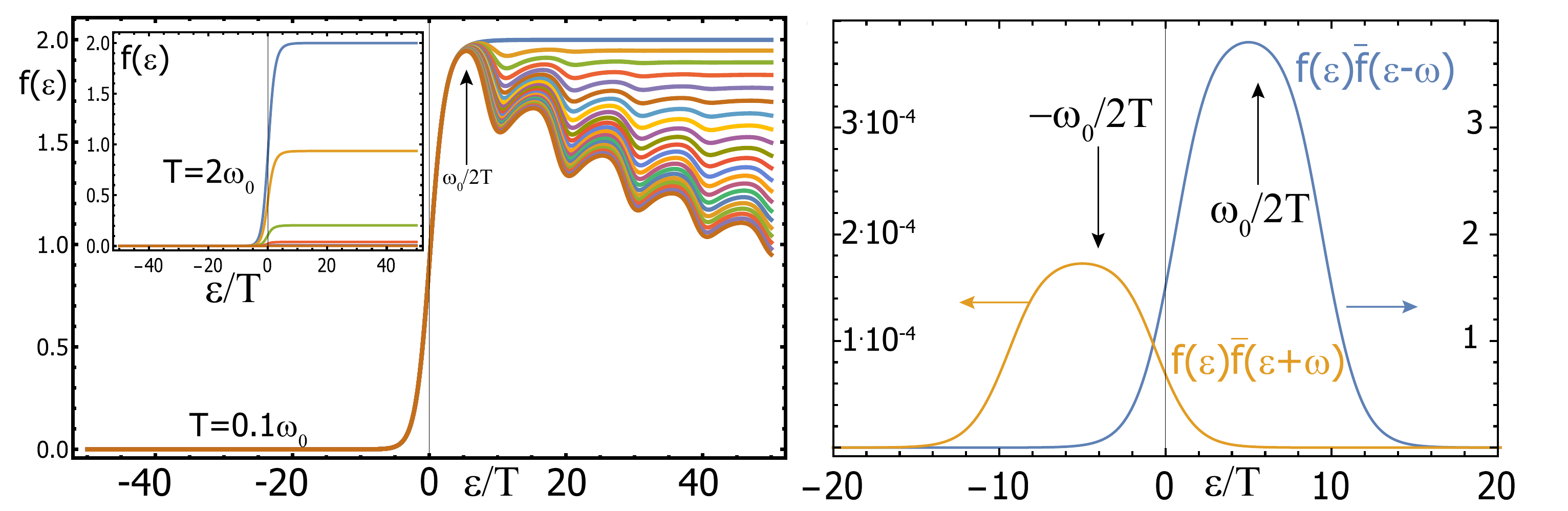}\centering

\caption{Left panel, main figure: low temperature evolution of non-diagonal
parameter $f$ in the low temperature case, $T=0.1\omega_{0}$ for
discrete times $t/\tau=0.2,\,0.4,\,\ldots4.0$. The inset: similar
evolution at high temperatures, $T=2\omega_{0}$, for $t/\tau=0.2,\,0.4,\,0.6,\,\ldots$
shows fast uniform decrease of the non-diagonal parameter. The right
panel displays the integrands of $I_{\pm}$ (\ref{eq:I_pm}) at $t/\tau=$2.0
that shows that $I_{\pm}$are dominated by narrow frequency ranges
around $\pm\omega_{0}/2$. This allow us to simplify the equations
by considering only the values of $f$ at these points. \label{fig:Low-temperature-evolution}}
\end{figure}

In order to describe the behavior of the solution in the quantum limit,
it is instructive to look at the results of the numerical solution
of Eq. (\ref{eq:phonon-instability-main}) at low temperatures, see
Fig. \ref{fig:Low-temperature-evolution}. In contrast to the classical
case, the function $f(\epsilon)$ does not preserve the shape with
the time evolution. However, one observes that the behavior does not
change at negative energies. At positive energies we observe a sequence
of peaks at energies $\epsilon_{n}=(n+1/2)\omega_{0}$ that are similar
to the first peak at $n=0$. This behavior can be qualitative understood
as follows.

At low temperatures Eq. (\ref{eq:I_pm}) implies that $I_{-}\gg I_{+}$
(see also Fig. \ref{fig:Low-temperature-evolution}b). If the terms
proportional to $I_{+}$ are dropped, Eq. (\ref{eq:phonon-instability-main}a)
describes the drift of $f(\epsilon)$ to high energies together with
relaxation. Similarly, Eq. (\ref{eq:phonon-instability-main}b) describes
the drift to negative energies and relaxation. The $L$-terms in Eqs.
(\ref{eq:phonon-instability-main}) lead to decay, so the frequency
regime where these terms dominate cannot contribute to the instability.
For frequencies $\left|\epsilon\right|>\omega_{0}$ the $L$-term
in Eq. (\ref{eq:phonon-instability-main}) is large so the region
responsible for the instability is $\left|\epsilon\right|<\omega_{0}$.
In the absence of $I_{+}$, the advection term proportional to $I_{-}$
removes perturbations from this region. Because the values of $f(\epsilon)$
at high energies does not feedback on low energies, the instability
disappears. In order to see the instability it is therefore essential
to keep the $I_{+}$ term in the region $\left|\epsilon\right|<\omega_{0}$.
From Fig. \ref{fig:Low-temperature-evolution}b we see that the main
contribution to $I_{\pm}$ comes from the vicinity of the frequencies
$\pm\omega_{0}/2$, so the integrals $I_{\pm}$ can be approximated
by $I_{-}\approx f_{+}^{2}$ and $I_{+}\approx f_{-}^{2}$ where $f_{\pm}=f(\pm\omega_{0}/2)$
(we drop non-essential numerical factors). For this reason we focus
on the time dependence of these two values of $f(\epsilon)$. In the
equation for $df_{-}/dt$ we can neglect the contribution of $I_{+}f(-3\omega_{0}/2)$
term because $f(-3\omega_{0}/2)\sim\exp(-\omega_{0}/T)f_{-}$. In
the equation for $df_{-}/dt$ we can neglect the contribution of $I_{-}f(3\omega_{0}/2)$
because it is proportional to $I_{-}\ll1$. Then the integral equations
(\ref{eq:phonon-instability-main}) reduce to two ordinary differential
equations: 
\begin{flalign}
\tau\frac{df_{+}}{dt} & =-L_{0}f_{+}+f_{+}^{2}f_{-}\label{eq:df+/dt}\\
\tau\frac{df_{-}}{dt} & =-L_{0}f_{-}+f_{-}^{2}f_{+}\label{eq:df-/dt}
\end{flalign}
where $L_{0}=2\exp(-\omega_{0}/2T)$. At low temperatures the linear
term in these equations becomes exponentially small, as a result the
instability developes exponentially slowly. Solving for the product
$f_{+}f_{-}$we get\begin{subequations} 
\begin{equation}
f_{+}f_{-}=\frac{L_{0}}{1+\exp\left[(t-t_{d})/t_{*}^{qu}\right]},
\end{equation}
and 
\begin{align}
f_{+}+f_{-} & =f_{0}\left(\frac{1}{1+\exp\left[(t-t_{d})/t_{*}^{qu}\right]}\right)^{1/2},
\end{align}
\label{f_pm_solution}\end{subequations}where $t_{*}^{qu}=\tau/(2L_{0})$.
The Eqs. (\ref{f_pm_solution}) should be compared with the solution
(\ref{eq:phi(t)}), we see that they describe similar relaxation but
with exponentially smaller rates. Although the behavior of the solution
(\ref{f_pm_solution}) is similar to the one in the high temperature
limit, there is an important difference: the relaxation is determined
only by a narrow advection region at low energies, whereas the high
energy region plays a passive role of a sink. Furthermore, the non-linearity
appears first at high energies but it does not affect the fact that
the dynamics is determined by the narrow region at low energies that
determines the value of $I_{+}.$ 

\subsection{Instability for electron-electron interaction. }

As for electron-phonon scattering one can focus only on the energy
dependence of the off-diagonal functions
\begin{flalign*}
\mathcal{F}_{ud}(\epsilon,p;t) & =f(\epsilon,t),\ \mathcal{F}_{du}(\epsilon,p;t)=-\bar{f}(\epsilon,t),
\end{flalign*}
 and assume that the diagonal functions correspond to the equilibrium.
As a result, the time evolution of the functions $f,\bar{f}$ is described
by the equations similar to Eq. (\ref{eq:phonon-instability-main}):

\begin{eqnarray}
\tau_{FL}\frac{\partial f}{\partial t} & = & -L_{F}\left(\frac{\epsilon}{T}\right)f+K(\epsilon,f,\bar{f}),\nonumber \\
\tau_{FL}\frac{\partial\bar{f}}{\partial t} & = & -L_{F}\left(\frac{\epsilon}{T}\right)\bar{f}+\bar{K}(\epsilon,f,\bar{f}),\label{eq:df/dt_el-el}\\
L_{F}(x) & = & 1+x^{2}/\pi^{2},\nonumber 
\end{eqnarray}
where 
\begin{equation}
\tau_{FL}^{-1}\sim T^{2}/E_{F}^{*}\label{eq:tau_FL}
\end{equation}
is the Fermi liquid relaxation rate, $E_{F}^{*}$ is the parameter
built from the electron density of states and the interaction constant.\footnote{In three dimensional Fermi liquid $E_{F}^{*}\sim E_{F}$ where $E_{F}$
is the Fermi energy, in two dimensional Fermi liquid it contains additional
$\ln(E_{F}/T)$ factors\cite{Aleiner2002} while in 1D models two
particle collisions do not lead to dissipation.} The functional $K(\epsilon,f,\bar{f})$ is of the second order in
$f$ and of the first order in $\bar{f}$:\begin{subequations} 
\begin{flalign}
K(\epsilon,f,\bar{f}) & =\int_{0}^{\infty}I_{-}(\omega)f(\epsilon-\omega)\frac{d\omega}{2\pi T}+\int_{0}^{\infty}I_{+}(\omega)f(\epsilon+\omega)\frac{d\omega}{2\pi T},\\
\bar{K}(\epsilon,f,\bar{f}) & =\int_{0}^{\infty}I_{+}(\omega)\bar{f}(\epsilon-\omega)\frac{d\omega}{2\pi T}+\int_{0}^{\infty}I_{-}(\omega)\bar{f}(\epsilon+\omega)\frac{d\omega}{2\pi T},\\
I_{\pm} & (\omega)=2\int\frac{d\epsilon}{2\pi T}f(\epsilon)\bar{f}(\epsilon\pm\omega).
\end{flalign}
\label{eq:K(f,fbar)}\end{subequations}

Formally these equations are similar to Eqs. (\ref{eq:phonon-instability-main})
for the electron-phonon scattering, the only difference is that instead
of one mode Eqs. (\ref{eq:K(f,fbar)}) contain the integral over frequencies.
At large $\epsilon\gg T$ the functionals $K(f,\bar{f})$ and $\bar{K}(f,\bar{f})$
are dominated by terms proportional to $I_{-}(\omega)$ that describes
drift to larger frequencies for $f(\epsilon)$ and to smaller frequencies
for $\bar{f}(\epsilon)$. The feedback that results in dissipation
is due to the energies $\epsilon\lesssim T$. The qualitative properties
of these equations are thus captured by the simplified equations for
two characteristic values of $f_{\pm}=f(\pm T)$. These equations
have exactly the same form as Eqs. (\ref{eq:df+/dt},\ref{eq:df-/dt}),
with the important difference that $L_{0}\sim1$. Thus their solution
is given by the equations (\ref{eq:phi(t)}) with characteristic decay
time $t_{*}\sim\tau_{FL}.$

It is very important that although the relaxation rate in a Fermi
liquid becomes very large at high energies, the processes involving
high energy electrons do not contribute to the instability of the
equations (\ref{eq:df/dt_el-el}) for the non-diagonal parts. Instead,
the instability is controlled by the same processes as the physical
relaxation and has characteristic time scale of the electron-electron
relaxation time at temperature $T$.

\section{Equations for spatial structure of the instability\label{sec:Spatial-structure-of}}

As we have seen in section \ref{sec:Instability-of-the} the instability
in zero dimensional models is always controlled by equations similar
to (\ref{eq:df/dt_el-ph}). This equation can be derived and solved
analytically in the case of electron-phonon model at high temperatures
but it provides the qualitative description in other cases as well.
To resolve the spatial structure of the instability we thus begin
with the electron-phonon model at high temperatures. The presence
of the spacial structure changes the quantum kinetic equations (\ref{eq:df/dt_el-ph})
very little (apart from introducing the spacial dependence). Because
the phonons in this model are local, the equations for $\theta$ and
$\bar{\theta}$ contain the fermionic functions taken at the same
spatial point. As in section \ref{subsec:Instability-in-electron-phonon}
in the limit of low phonon density ($\eta\ll1$) the phonon relaxation
is fast, so we can solve for local $\theta$ and $\bar{\theta}$:

\begin{flalign*}
\theta(r) & =\frac{1}{2\omega_{0}}\int d\epsilon f(\epsilon,r)\bar{f}(\epsilon-\omega_{0},r),\\
\bar{\theta}(r) & =\frac{1}{2\omega_{0}}\int d\epsilon f(\epsilon,r)\bar{f}(\epsilon+\omega_{0},r),
\end{flalign*}

Performing the standard spatial gradient expansion in the LHS of the
kinetic equation (\ref{eq:KinEq_El_LHS}), we find that $df/dt$ in
(\ref{eq:df/dt_el-ph}) acquires an additional diffusion term. Parametrizing
the solution by the ansatz (\ref{eq:f(epsilon)}) we obtain the final
equation

\begin{equation}
\frac{\partial\phi}{\partial t}-D_{*}\nabla^{2}\phi=-\frac{2\left(\phi-\phi^{3}\right)}{t_{*}}.\label{eq:dphi/dt_General_0}
\end{equation}

This equation is the central result of this paper. As we argue below
it holds (with small modifications) for other models as well.

At non-zero temperature the electron-phonon interaction leads to the
diffusive motion of electrons characterized by the momentum relaxation
time so that the diffusion coefficient $D_{*}=v_{F}^{2}\tau_{tr}/d$,
where $v_{F}$ is the Fermi velocity and $d$ is the spatial dimensionality.
At high temperatures $\omega_{0}\ll T$ the transport relaxation rate
is given by $1/\tau_{tr}=\lambda^{2}\nu(n_{ph}T/\hbar\omega_{0})$,
with $(n_{ph}T/\hbar\omega_{0})_{ph}$ having the meaning of the thermal
phonon density. In this case the energy relaxation of the electrons
becomes parametrically slower than its momentum relaxation: $1/\tau_{e}=(\omega_{0}/T)\lambda^{2}\nu n_{ph}$.

The diffusion approximation used to derive Eq. (\ref{eq:dphi/dt_General_0})
can be rigorously justified only if the resulting gradient of $\phi$
is small on the scale of the mean free path, $v_{F}\tau_{tr}$. This
happens only if $t_{*}^{cl}\gg\tau_{tr}$ which can occur if the electron
diffusion is additionally slowed down by the impurity scattering $1/\tau_{tr}=1/\tau_{tr}^{(ph)}+1/\tau_{tr}^{(imp)}\gg1/\tau_{tr}^{(ph)}$.

Both the diffusion coefficient $D_{*}$ and the time $t_{*}^{cl}$
depend on the local temperature $T(r,t)$ and the electron density
$n(r,t)$. Those quantities are described by the standard diffusion
and thermal diffusion equations for the diagonal components and their
solutions has to be used as entry parameters for Eq. (\ref{eq:dphi/dt_General_0}).
This scheme gives the complete description of the quantum butterfly
effect. Notice that depending on the particular model, $D_{*}$ may
coincide with the particle or thermal diffusion coefficients or may
be different from those by a numerical factor.

A very similar equation can be put forward for the model of the electron-electron
interaction. Taking into account that the effective equations for
electron-electron interaction is formally the same as that for electron-phonon
case, we write

\begin{equation}
\left(\frac{\partial}{\partial t}+v\nabla\right)\phi-D_{*}\nabla^{2}\phi=-\frac{2\left(\phi-\phi^{3}\right)}{t_{*}}.\label{eq:dphi/dt_General_1}
\end{equation}
The only modification here is the appearance of the drift term $v\nabla$which
is dictated by the Galilean invariance for $\xi_{p}=p^{2}/2m-\epsilon_{F}$.
The macroscopic velocity $v(r,t),$ the local temperature $T(r,t)$,
and the electron density $n(r,t)$ are controlled by the usual equations
of local hydrodynamics and thermal (entropy) diffusion \cite{LandauLifshitzv10}.
It is possible to generalize Eq. (\ref{eq:dphi/dt_General_1}) for
the case of relativistic hydrodynamics. Based on Lorentz invariance
one obtains

\begin{equation}
\left(u^{i}\partial_{i}+D_{*}\partial_{i}\partial^{i}\right)\phi=-\frac{2\left(\phi-\phi^{3}\right)}{t_{*}},\label{eq:dphi/dt_General_2}
\end{equation}
where covariant and contravariant component are related by the arbitrary
metric tensor and $u^{i}$ is standard four component velocity vector
with local constrain $u_{i}u^{i}=1.$

To close the section, let us emphasize that the coefficients $D_{*,}t_{_{*}}$
do not affect the diagonal entropy production and do not enter the
usual Onsager relations. It is unknown to us whether there is an analogue
of the $H$-theorem that includes the non-diagonal distribution functions
as well.

\section{Spatial Propagation of the instability: combustion waves.\label{sec:Spatial-Propagation-of}}

The equations (\ref{eq:dphi/dt_General_0},\ref{eq:dphi/dt_General_1},\ref{eq:dphi/dt_General_2})
for the spatial structure of the instability are well known in the
theory of combustion. In particular, Eq. (\ref{eq:dphi/dt_General_0})
is very similar to Fisher- Kolmogorov\textendash Petrovsky\textendash Piscounov
equation (FKPP) \cite{Fisher1937,KPP} 
\[
\frac{dy}{dt}-\nabla^{2}y=y(1-y).
\]
All equations of this type possess two stationary solutions $y=0$
and $y=1$ in case of FKPP, one of them is stable, another is not.
In particulr, our Eq. (\ref{eq:dphi/dt_General_0}), displays the
instability of the solution $\phi(r)=1$ that evolves according to
the following scenario. After being seeded at time $t=0$ with the
small deviation $\delta\phi(r)=1-\phi(r)\ll1$, in a region around
$0$ (i.e. $\delta\phi=0$ for $r>R_{c}$) the instability remains
localized in the area where it was seeded ($r<R_{c})$ for the time
$t_{d}\sim\ln(1/\delta\phi)$. After this initial period, the instability
starts to grow spatially forming a non-linear wave that moves with
a well defined velocity $v_{cw}$.

For Eq. (\ref{eq:dphi/dt_General_0}) in 1D the solution $\phi_{f}(x-v_{cw}t)$
for the front moving with constant velocity $v$ obeys the equation
\begin{equation}
t_{*}\left(v_{cw}\frac{d\phi_{f}}{dx}+D_{*}\frac{d^{2}\phi_{f}}{dx^{2}}\right)=2\phi_{f}(1-\phi_{f}^{2})\label{eq:FrontForm}
\end{equation}
As is established in the theory of combustion \cite{Zeldovich1985},
the value of the front velocity can be found from the study of the
solution of Eq.(\ref{eq:FrontForm}) at $x\rightarrow\infty$ where
$\delta\phi\rightarrow0$. At $\delta\phi\ll1$ the solution of Eq.
(\ref{eq:FrontForm}) behaves as $\delta\phi\sim\exp(-kx)$ with $k$
that is real at 
\begin{equation}
v_{cw}^{2}\geq16D_{*}/t_{*}.\label{eq:v_stability}
\end{equation}
For the initial conditions that correspond to $\delta\phi=0$ for
$r>R_{c}$ the solution quickly converges to the one moving with the
minimal velocity allowed by the constraint (\ref{eq:v_stability}).
The presence of other solutions (with higher velocities) is due to
the fact that for the (non-physical) initial conditions that differ
from unity everywhere, the instability develops at large $r$ might
develop independently of the seed at small $r$. One concludes that
the combustion wave moves with velocity 
\[
v_{cw}=4\sqrt{D_{*}/t_{*}}
\]

Note that for electron-phonon and electron-electron models $D_{*}\sim t_{*}v_{F}^{2}$
in the absence of electron-impurity and elastic scattering, so that
the front velocity $v_{cw}\sim v_{F}$. Because no perturbation (even
unphysical one) can propagate with velocity larger than $v_{F}$,
$v_{cw}\lesssim v_{F}.$

In order to check the conclusions of the semi-quantitative analysis
presented above we have studied numerically the front propagation
in the dimensionless equation 
\begin{equation}
\frac{d\phi}{dt}=\nabla^{2}\phi+2\phi(\phi^{2}-1)\label{eq:dphi/dt_CW1}
\end{equation}
and in the similar equation describing evolution of both electrons
and phonons\begin{subequations} 
\begin{eqnarray}
\frac{d\phi}{dt} & = & \nabla^{2}\phi+2(\Theta-1)\phi,\\
\frac{d\Theta}{dt} & = & \phi^{2}-\Theta
\end{eqnarray}
\label{eq:dphi/dt_CW2}\end{subequations}that describes the situation
in which the phonon dynamics is of the same order as electron one
(i.e. $\eta=1$). We found that in all cases and in all dimensions
($d=1,2,3$) the front quickly assumes a well defined shape and start
to move with the constant velocity. We note that this conclusion for
the two component (electron and phonon) systems is not obvious because
such equations are known to display more complex behavior in some
cases.

\begin{figure}[h]
\includegraphics[width=0.9\textwidth]{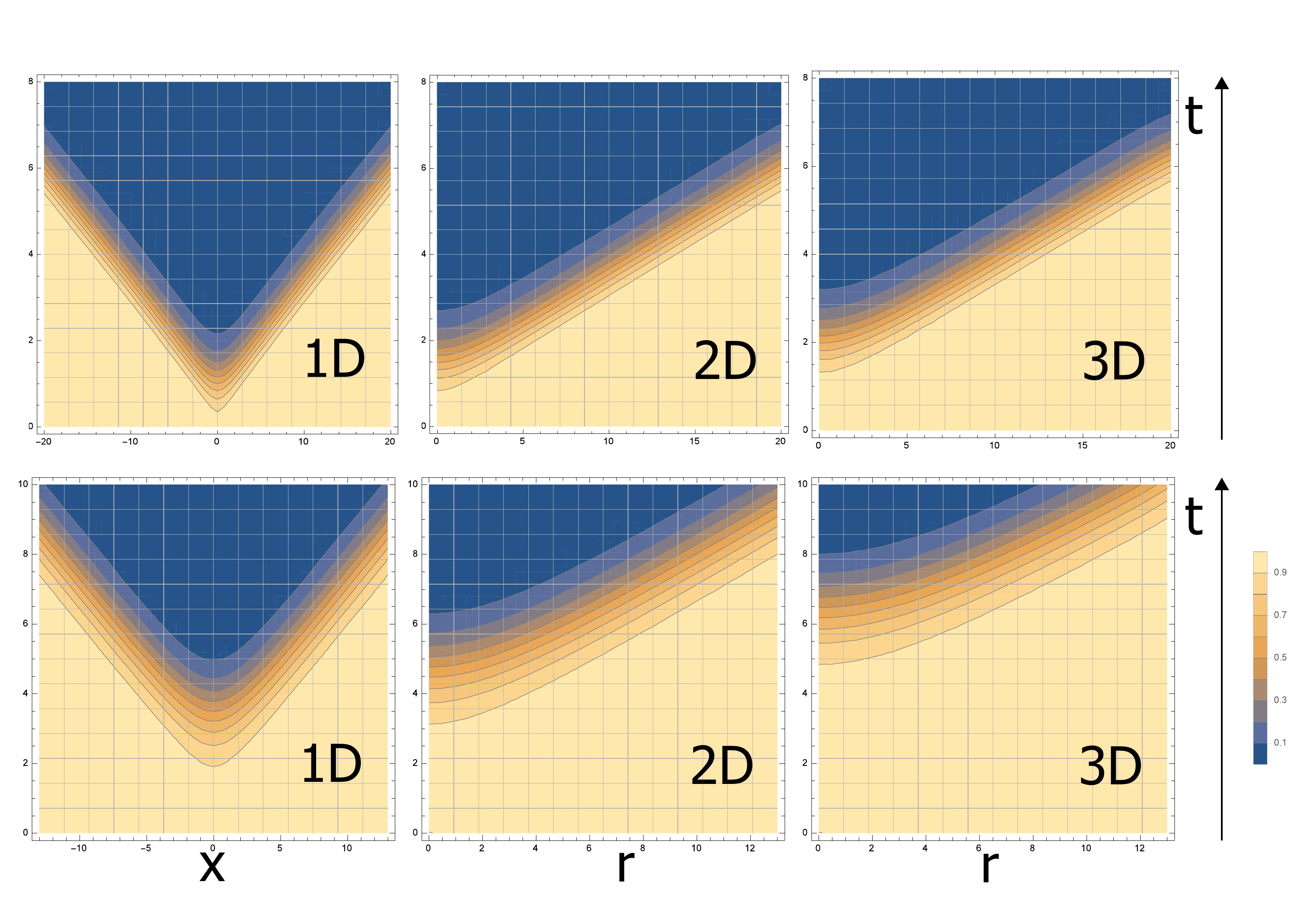}\centering

\caption{Propagation of instability in different dimensions. The upper panel
shows the solution of (\ref{eq:dphi/dt_CW1}) for one component electron
system, the lower for the combined system of electron and phonons
(\ref{eq:dphi/dt_CW2}). \label{fig:Propagation-of-instability}}
\end{figure}

\section{Time and spatial dependence of out-of-time-ordered correlators. \label{sec:Initial-conditions-for}}

We now apply the findings of the previous sections, namely, the instability
of the off-diagonal part of the Green function to the computation
of the out-of-time-ordered correlator (\ref{eq:A_rhorho}). In the
conventional theory the correlator of two operators at large separations
in time or space factorizes 
\begin{equation}
\left\langle \hat{\mathcal{S}}_{0}\hat{\mathcal{R}}_{t,t'}\right\rangle =\left\langle \hat{\mathcal{S}}_{0}\right\rangle \left\langle \hat{\mathcal{R}}_{t,t'}\right\rangle \label{eq:factorization}
\end{equation}
The corrections to this factorization are given by irreducible correlator
that decreases quickly with distance and time. In the electron models
considered here the irreducible part is small in $1/p_{F}r$ and $1/\epsilon_{F}t$.
Furthermore, in a conventional theory one can evaluate both averages
in the RHS of (\ref{eq:factorization}) against the background of
the unperturbed states.

The crucial difference of the two Worlds theory is that the second
term in this factorization is unstable. Thus, it is not correct to
replace it by its value for the fully correlated, unperturbed state:
a small deviation from this value at short distances grows quickly
and eventually reduces it to zero in the whole system. Instead one
should use for it the results of the solution of the equations for
the Green functions discussed in previous sections. In particular,
for the response operator in correlator (\ref{eq:A_rhorho}) we get
\begin{equation}
\left\langle \hat{\mathcal{R}}_{t,t'}(r)\right\rangle =2\pi\nu\left[f(\epsilon,t,r)-\bar{f}(\epsilon,t,r)\right].\label{eq:R_tt'_final}
\end{equation}
where we emphasized that the augmented distribution functions $f$
and $\bar{f}$ are generally the functions of the position in the
space as well. The space-time dependence of these functions is determined
by the equations derived in Sections \ref{sec:Instability-of-the}-\ref{sec:Spatial-Propagation-of}.

The average of the source operator is a constant factor, for the correlator
(\ref{eq:A_rhorho}) it is given by the total density of electrons:
\begin{equation}
\left\langle \hat{\mathcal{S}}_{0}\right\rangle =n_{el}.\label{eq:S_0_aver}
\end{equation}

As discussed in Section \ref{sec:Instability-of-the} the time dependence
of the augmented distribution functions is simplified in the high
temperature regime of the electron-phonon model. In this case the
form of the energy dependence of the augmented distribution function
does not change with time, the time dependence shows up only in the
factor $\phi(r,t)$: $f(\epsilon,t,r)=\phi(r,t)f_{0}(\epsilon)$.
In this case we can write the final result for the Wigner transform
of the out-of-time ordered correlator in the closed form 
\begin{align}
\mathcal{\tilde{A}}_{\rho\rho}(\epsilon,t,r) & =2\pi\nu n_{0}(\epsilon)n_{el}\phi(t,r),\label{eq:A_rhorho_semifinal}\\
\mathcal{A}_{\rho\rho}(t',t,r) & =\int(d\epsilon)e^{-i\epsilon(t'-t)}\mathcal{\tilde{A}}_{\rho\rho}(\epsilon,\frac{t+t'}{2},r),
\end{align}
Here $\phi(t,r)$ is the solution of the equations (\ref{eq:dphi/dt_General_0}-\ref{eq:dphi/dt_General_2})
appropriate for a particular model with the initial conditions 
\begin{equation}
\phi(0,r)=1-\delta\phi(r)\label{eq:phi(0)}
\end{equation}
Here 
\begin{equation}
\delta\phi(r)\simeq\tilde{\delta}(r)/n_{el}\simeq\tilde{\delta}(r)p_{F}^{_{-d}}\label{eq:delta_phi}
\end{equation}
describes the perturbation resulting from the introduction of one
extra electron in down World, which serves as a seed of the instability.
Here $\tilde{\delta}(r)$ denotes the smeared $\delta$-function that
appears because the equations for the distribution function are valid
only at the time scales larger than collision time $\tau_{tr}$, so
the addition of one particle at time $t=0$ in the down World translates
in the density spead over distance $l_{tr}\sim v_{F}\tau_{tr}$ for
the initial conditions of the Eqs. (\ref{eq:dphi/dt_General_0}-\ref{eq:dphi/dt_General_2}).
As a result the $\delta-$function in Eq. (\ref{eq:delta_phi}) has
to be replaced by $\tilde{\delta}(r)$ which is smeared at the distances
of mean free path, $l_{tr}\sim v_{F}\tau_{tr}$ . 

The solutions of the equations (\ref{eq:dphi/dt_General_0}-\ref{eq:dphi/dt_General_2})
correspond to the propagation of the front as illustrated by Fig.
\ref{fig:Propagation-of-instability}.

Note that the particular symmetric form (\ref{eq:R_tt'}) of the response
operator computed here has the property that it vanishes at coinciding
times and coordinates. This property disappears for less symmetric
form of the correlators, for instance if $\tau_{1}^{a}$ is replaced
by, e.g. $\tau_{-}=\frac{1}{2}(\tau_{1}-i\tau_{2})$, in the definition
of the response operator (\ref{eq:R_tt'}). 
\begin{equation}
\mathcal{\tilde{A}}'_{\rho\rho}(t,r)=\left\langle T_{\mathcal{C}}\left(\boldsymbol{\bar{\Psi}}(t',r)(\tau_{1}^{K}\otimes\tau_{-}^{a})\boldsymbol{\Psi}(t,r)\right)\mathcal{\hat{S}}_{0}\right\rangle \label{eq:A'_def}
\end{equation}
In this case the first term in (\ref{eq:R_tt'_final}) disappears
and we get after integration over energies 
\begin{equation}
\mathcal{\tilde{A}}'_{\rho\rho}(t,r)=2n_{el}^{2}\phi(t,r).\label{eq:A'_rhorho_final}
\end{equation}

At low temperatures for electron-phonon model and for electron-electron
interaction at any temperature the energy dependence of the augmented
distribution function changes with time as well (Sections \ref{subsec:Quantum-limit-()},
\ref{subsec:Electron-electron-interaction}). In this case the equations
for the augmented distribution function are more complicated but the
solution remains qualitatively similar.

In all cases, the augmented distribution function that controls the
spacial and time dependence of the out-of-time-ordered correlator
describes the front propagation, the state of the system before the
front has not been affected yet by perturbation whilst the state of
the system behind the front is characterized by exponentially vanishing
correlations: 
\begin{equation}
\mathcal{\tilde{A}}_{\rho\rho}(\epsilon,t,r)=2\pi\nu n_{0}(\epsilon)n_{el}\begin{cases}
\exp\left(-\frac{t-t_{d}-r/v_{cw}}{2t_{*}}\right) & t>r/v_{cw}+t_{d}\\
1-\exp\left(\frac{t-t_{d}-r/v_{cw}}{t_{*}}\right) & t<r/v_{cw}+t_{d}
\end{cases}\label{eq:A_rhorho_final}
\end{equation}

The delay time $t_{d}$ in these equations is controlled by the initial
conditions (\ref{eq:phi(0)}) to the Eqs. (\ref{eq:dphi/dt_General_0}-\ref{eq:dphi/dt_General_2})
or similar. It depends only logarithmically on the strength of the
initial perturbation: 
\begin{equation}
t_{d}=t_{*}\left|\ln\left[\delta\phi(0)\right]\right|\label{eq:t_d_1}
\end{equation}

Equation (\ref{eq:delta_phi}) enables us to estimate the strength
of the initial perturbation. Indeed, $\delta(0)\simeq1/(v_{F}\tau_{tr})^{d}$
. Thus, we estimate $\delta\phi(0)\simeq1/(p_{F}l_{tr})^{d}$ and
the delay time 
\begin{equation}
t_{d}=t_{*}d\ln(p_{F}v_{F}\tau_{tr})\label{eq:t_d_2}
\end{equation}
It is worthwhile to notice that this expression is somewhat similar
to the Ehrenfest time \cite{Aleiner1996,Aleiner1997} appearing as
the delay time for the quantum correction in quantum chaos for non-interacting
system. In this one electron problem the real instability does not
occur.

In a finite size system of spatial size $R$ the correlator (\ref{eq:A_rhorho_final})
decreases exponentially to zero for all $r<R$ after $t_{scr}=t_{d}+R/v_{cw}$.
The time $t_{scr}$ has the meaning of the time at which the two worlds
become completely uncorrelated due to a local perturbation, this is
also the time that it takes for the quantum information to be spread
over the whole system (scrambling time). We see that although the
propagation of the information is controlled by diffusion, it occurs
with a constant velocity due to the non-linearity of the equations.
The diffusion coefficient controls the velocity of this propagation.

The propagation with constant velocity (\ref{eq:A_rhorho_semifinal},\ref{eq:A_rhorho_final})
also indicates that in a chaotic many body system the entanglement
entropy spreads ballistically despite the diffusive nature of the
dynamics. This analytical result confirms the empirical conclusions
reached in a number of numerical works.

As we discussed above, the conclusions of the linear propagation of
the quantum butterfly effect controlled by the combustion equations
is quite general. The details of the equations are sensitive to the
microscopic model but the linear propagation similar to combustion
front occurs in all of them.

\section{Discussion and conclusions\label{sec:Discussion-and-conclusions} }

We developed the technique to study the out-of-time-ordered correlators,
such as Eq. (\ref{eq:A_general}), based on the extension of Keldysh
technique. Similarly to standard Keldysh technique, the augmented
technique enables the analytical study of systems in different limits,
in particular to obtain the leading result in the quasiclassical approximation
and systematic corrections to it. As well as in the Keldysh technique
the quasiclassical approximation is valid provided that the particle
motion between collisions is quasiclassical whilst collision themselves
can be quantum.

We limited ourselves to the leading quasiclassical terms that result
in the equations similar to the kinetic equation in traditional statistical
mechanics. We found that they describe all (or most of all) non-trivial
behavior of the out-of-time-ordered correlators. The major difference
from the traditional kinetic equation is the appearance of the off-diagonal
functions, superficially similar to the conventional distribution
function. However, unlike the state occupation probabilities, these
new functions also describe the overlap between two copies of the
system. The kinetic equation for the off-diagonal functions is dramatically
different from that for the diagonal functions: the outgoing term
depends on both diagonal and off-diagonal functions whilst the incoming
term contains only off-diagonal ones.

The solution with initially unit overlap between two copies becomes
unstable when disturbed by a very small perturbation, the phenomenon
known as quantum butterfly effect. This instability is described at
long times (longer than collision times) by non-linear diffusion equations
similar to those appearing in the combustion front propagation. After
an initial transient behavior the front of the propagating wave acquires
a constant velocity and a shape that does not depend on the initial
conditions (Section \ref{sec:Spatial-Propagation-of}). In the electron
models studied in this paper, the velocity of the front is of the
order (but less than) the Fermi velocity that serves as natural bound
for the propagation speed. In the presence of impurity scattering
the velocity of the front can become parametrically slower than Fermi
velocity. The microscopic model of electrons interacting with the
dilute set of oscillators solved in this work might provide the description
of the loss of coherence in the set of two level systems (TLS) that
provide both elastic and non-elastic scattering for electrons with
the latter becoming small at low temperatures.

Our work suggests a number of exciting developments. First, the quantum
butterfly effect studied here can be viewed as the result of the gradual
entanglement of the local degrees of freedom with larger and larger
part of the surrounding system, and thus is likely to be related to
the propagation of entanglement entropy discussed extensively recently
\cite{Calabrese2005,Hartman2013,Casini2015,bardarson2012,Roberts2015}.
Our results would enable us to put these works on the firm ground
of an analytical theory if the relation between non-diagonal correlators
and entanglement entropy is established. We hope to return to this
point in future works.

The quantum butterfly effect can be studied numerically and compared
with the analytical theory developed here. Also, the destruction of
the coherence between two copies of the system might be a useful tool
to study the appearance of the arrow of time in the systems described
by the unitary evolution. Finally, the destruction of quantum coherence
between two copies of the system is a very important phenomenon for
the quantum information protocols that are based on the construction
of the initially perfectly entangled states of two (or more) interacting
qubit systems because small perturbation to one of these systems would
result in a spreading decoherence wave described by our equations.

The spatial and time scales of the effective non-linear diffusive
equations that describe the instability of the coherent solution are
sensitive to the details of the microscopic theory. Furthermore, their
relation to the ones appearing in physical observables is not expected
to be universal. Thus they might provide a new tool and the new way
of thinking about microscopically different systems that display similar
properties such as conductivity.

Our formalism can be extended to the study of many body localization
by augmenting the formalism developed in the work \cite{Basko2006}.
This would provide the analytical approach and qualitative understanding
to the problem for which only numerical results are currently available.\cite{Chen2016,Shen2016,Fan2016,Huang2016}
It might even help to describe the transition itself and even the
entanglement propagation in generic glassy systems. Moreover, the
question of the propagation of the decoherence front in localized
systems is similar to the problem of the decoherence propagation in
integrable systems. Note that according to \cite{Basko2006,Glazman2007a}
in localized and integrable systems the collision integral disappears
resulting in the suppression of the chaotic behavior that is responsible
for the quantum butterfly effect.

Finally, the microscopic systems studied in this work are described
by the combustion equations that display only laminar solution. However,
combustion equations for systems with a few components are known to
display a large variety of interesting behaviors: Turing instabilities \cite{Turing1952},
Zhabotinsky cycles \cite{Zhabotinsky1964} to name just a few. It
remains to be seen if these solutions are realized in microscopic
models as the instabilities of the correlated worlds solution. In
particular, they might appear as the solutions against the background
of non-equilibrium states such as turbulent hydrodynamics of normal
or superfluid liquid.

\section{Acknowledgement}

We acknowledge extremely useful discussions with Alexei Kitaev and
the hospitality of CTP CSIBS, Daejeon, Korea. Our research was supported
by ARO grant W911NF-13-1-0431 and by the Russian Science Foundation
grant \# 14-42-00044.

\bibliographystyle{elsarticle-num}
\bibliography{qip}

\end{document}